\documentclass[10pt]{article}
\usepackage{rhstyle}
\usepackage{soul}
\hypersetup{pdftitle={SecProbe: Adaptive Evaluation of Coding Agents on Cybersecurity Vulnerabilities}, pdfauthor={Xiaonan Luo, Yue Huang, Kehan Guo, Ping He, Chuan Zou, Chujie Gao, Lichi Li, Yuchen Ma, Zhangchen Xu, Zichen Chen, Yufei Han, Xiangliang Zhang}}

\renewcommand{\headrulewidth}{0pt}
\fancypagestyle{firstpage}{%
  \fancyhf{}%
  \renewcommand{\headrulewidth}{0pt}%
  \fancyfoot[C]{\footnotesize\color{BakeLightGray}\thepage}%
}

\newcommand\cmark{\textcolor{green}{\ensuremath{\checkmark}}}
\newcommand\xmark{\textcolor{muted!70}{\ensuremath{\times}}}
\colorlet{oursrow}{accent!10}
\newcommand\tgroup[2]{\multicolumn{#1}{@{}l}{\sffamily\bfseries\scriptsize\color{accentdk}\MakeUppercase{#2}}}

\definecolor{pbIndigo}{HTML}{3557D5}
\definecolor{pbAmber}{HTML}{E8890C}
\definecolor{pbTeal}{HTML}{12A594}
\definecolor{pbViolet}{HTML}{7C5CD6}
\definecolor{pbRose}{HTML}{D6457A}
\definecolor{pbInk}{HTML}{202124}
\newcommand\pbaccent[2]{\expandafter\def\csname pbaccent@#1\endcsname{#2}}
\pbaccent{Online Security Researcher}{pbIndigo}
\pbaccent{Task Architect}{pbAmber}
\pbaccent{Exploit Test Strategist}{pbRose}
\pbaccent{Repository Engineer}{pbTeal}
\pbaccent{Security Reviewer}{pbViolet}
\newcommand\pbpick[1]{%
  \ifcsname pbaccent@#1\endcsname
    \colorlet{pbAccent}{\csname pbaccent@#1\endcsname}%
  \else\colorlet{pbAccent}{pbIndigo}\fi}

\newcommand\pblabel[1]{\par\addvspace{3pt}{\sffamily\bfseries\scriptsize\color{pbAccent!80!black}\MakeUppercase{#1}}}

\newtcolorbox{promptbox}[2][]{
  enhanced,
  code={\pbpick{#2}},
  width=\linewidth,
  sharp corners=west, arc=4pt,
  boxrule=0pt, frame hidden,
  colback=pbAccent!4!white,
  coltext=pbInk,
  borderline west={2.5pt}{0pt}{pbAccent},
  left=11pt, right=11pt, top=5pt, bottom=9pt,
  fontupper=\small,
  title={#2},
  fonttitle=\sffamily\bfseries\small,
  coltitle=pbAccent!80!black,
  colbacktitle=pbAccent!12!white,
  toptitle=4pt, bottomtitle=4pt, lefttitle=11pt,
  titlerule=0pt,
  before upper={\let\textbf\pblabel},
  #1
}

\title{SecProbe: Adaptive Evaluation of Coding Agents on Cybersecurity Vulnerabilities}
\newcommand{\afflogo}[2][0.95em]{\raisebox{-0.12em}{\includegraphics[height=#1]{figures/affiliation-logos/#2}}\hspace{0.22em}}
\author{%
  \makebox[\linewidth][l]{%
  Xiaonan Luo\textsuperscript{1,2,*}\hspace{0.8em}
  Yue Huang\textsuperscript{1,2,*}\hspace{0.8em}
  Kehan Guo\textsuperscript{1}\hspace{0.8em}
  Ping He\textsuperscript{3}\hspace{0.8em}
  Chuan Zou\textsuperscript{4}\hspace{0.8em}
  Chujie Gao\textsuperscript{1}}\\[3pt]
  \makebox[\linewidth][l]{%
  Lichi Li\textsuperscript{2}\hspace{0.8em}
  Yuchen Ma\textsuperscript{5}\hspace{0.8em}
  Zhangchen Xu\textsuperscript{2,6}\hspace{0.8em}
  Zichen Chen\textsuperscript{2,7}\hspace{0.8em}
  Yufei Han\textsuperscript{8}\hspace{0.8em}
  Xiangliang Zhang\textsuperscript{1}}\\[6pt]
  {\small\color{muted}%
  \textsuperscript{1}\afflogo{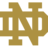}University of Notre Dame\quad
  \textsuperscript{2}\afflogo{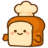}Bake AI\quad
  \textsuperscript{3}\afflogo{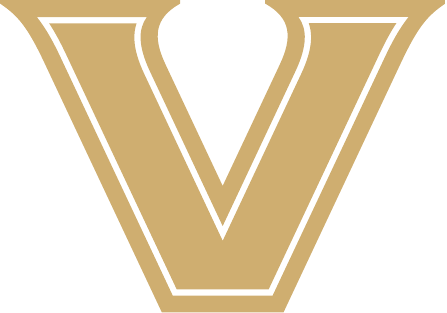}Vanderbilt University\\
  \textsuperscript{4}\afflogo{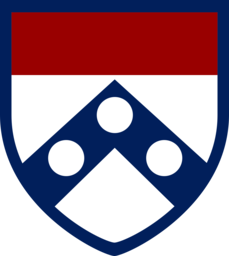}University of Pennsylvania\quad
  \textsuperscript{5}\afflogo{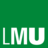}LMU Munich\quad
  \textsuperscript{6}\afflogo{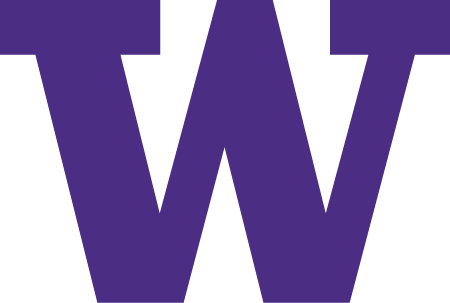}University of Washington\\
  \textsuperscript{7}\afflogo{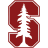}Stanford University\quad
  \textsuperscript{8}\afflogo[0.8em]{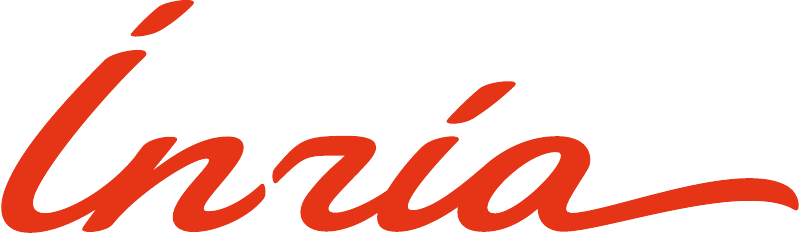}Inria\\[3pt]
  \textsuperscript{*}Equal contribution}}
\date{Preprint, \today}

\begin{document}
\maketitle
\thispagestyle{firstpage}

\begin{abstract}
\noindent
Assessing cybersecurity vulnerability awareness in coding agents requires evaluations that reveal capability gaps and remain informative as models evolve. Static benchmarks offer fixed coverage and difficulty, while scarce vulnerable repositories and costly expert authoring limit their renewal at scale. We introduce \textsc{SecProbe}, a framework for adaptive evaluation that combines Item Response Theory (IRT) with on-demand synthesis of repository-scale vulnerability-repair tasks. From observed performance, \textsc{SecProbe} estimates agent ability and identifies where additional evidence is most informative, selecting existing tasks or synthesizing new ones accordingly. As one use case, we construct 353 tasks spanning six programming languages and 151 CWE types and evaluate nine frontier models with two agent harnesses. Success rates peak at 28.33\%, highlighting substantial gaps in vulnerability recognition and repair. Compared with random and one-shot baselines, \textsc{SecProbe} achieves comparable agent ability estimates while requiring agents to solve up to 29.5\% fewer tasks. These results support adaptive evaluation as an efficient and discriminative approach to assessing cybersecurity vulnerability awareness.
\end{abstract}

\section{Introduction}
\label{intro}

Large language model (LLM)-based coding agents \citep{wang-etal-2025-swe,vergopoulos2025automated,wang2025swemirrorscalingissueresolvingdatasets} increasingly navigate large repositories, reason across files \citep{rashid2025swepolybenchmultilanguagebenchmarkrepository,zhang2026sweexplorebenchmarkingcodingagents,liu-etal-2025-codexgraph}, and implement complex changes using tools and tests \citep{wang-etal-2025-swe,lee2025secbench,deng2026swebench,ma2026swefficiency}. These capabilities amplify the consequences of insecure reasoning: functionally correct code can still contain vulnerabilities that agents fail to recognize or address \citep{kozak2025developeraidsecuritydebt,sajadi2025safeaigeneratedpatcheslargescale,chen-etal-2026-securevibebench,mou-etal-2025-really}. As agents assume greater responsibility for software development, a central question arises: \textbf{how can we efficiently and reliably evaluate their cybersecurity vulnerability awareness?} We assess this awareness through their ability to identify and repair vulnerabilities.

Existing evaluations typically use fixed security tasks drawn from real-world repositories \citep{wang-etal-2025-cve,lee2025secbench,chen-etal-2026-securevibebench} or authored by experts \citep{arx2026autobaxbuilder,spence2026challengeagenticcybersecurityrealistic,vergopoulos2025automated}. Although these benchmarks standardize comparison, they face three limitations. \textit{First}, assigning the same tasks to agents with different capabilities can spend substantial effort on tasks that are too easy, too difficult, or redundant to distinguish them, while leaving important capabilities underexplored. \textit{Second}, expanding coverage is costly: repositories with reproducible vulnerabilities, reliable patches, and security tests are scarce \citep{wang-etal-2025-cve,lee2025secbench,arx2026autobaxbuilder}, and authoring new tasks requires substantial expertise \citep{arx2026autobaxbuilder,spence2026challengeagenticcybersecurityrealistic,vergopoulos2025automated}. \textit{Third}, as agents improve and benchmark artifacts enter training corpora, fixed evaluations may become less reliable.

These limitations parallel a long-standing problem in assessment: estimating ability with a limited testing budget. Computerized adaptive testing (e.g., GRE, GMAT) uses previous responses to select subsequent items that provide additional information. Inspired by this principle and recent advances in Item Response Theory (IRT)-based benchmarking \citep{polo2024tinybenchmarks,ho2025rosettastoneaibenchmarks,truong2025reliable,lang-etal-2025-beyond}, we propose to frame cybersecurity evaluation as an \textbf{adaptive assessment process}: observed outcomes guide what to probe next. However, adaptive selection requires an adequate task bank, whereas existing security collections cover limited vulnerability families (detailed in Table~\ref{tab:benchmark_comparison}). The central obstacle is that \textbf{the informative tasks an assessment needs may not yet exist}; selection alone cannot fill these gaps.

To address this gap, we introduce \textsc{SecProbe}, a framework for \textbf{adaptive evaluation of coding-agent vulnerability awareness} that combines adaptive task selection with on-demand task synthesis. Agents are evaluated on repository-level codes that require them to \textit{detect and repair} vulnerabilities. Their performance is measured primarily by whether their repairs block exploitation samples designed for the injected vulnerabilities, with reference patches providing supplementary evidence of repair correctness. An IRT-based evaluation loop uses these outcomes to estimate agent ability, task difficulty, and discrimination, and determine which tasks would provide the most additional information. These estimates guide selection from the available task pool. When suitable tasks are missing, \textsc{SecProbe} coordinates specialized agents to synthesize new tasks conditioned on a target Common Weakness Enumeration (CWE) class and software context to meet the assessment needs. By making task construction responsive to evaluation outcomes, \textsc{SecProbe} concentrates its assessment budget on unresolved capability gaps and renews its task pool as agents evolve.

As one use case, we construct 353 tasks spanning six programming languages and 151 CWE types, assess their quality through quantitative analysis and expert review, and evaluate nine frontier models with two agent harnesses. \textsc{SecProbe} achieves comparable agent ability estimates while requiring agents to solve up to 29.5\% fewer tasks than baselines. The strongest evaluated configuration achieves only a 28.33\% success rate, highlighting persistent challenges in vulnerability identification and repair.

\textbf{\ul{Contributions.}} Our contributions are summarized below:
\begin{itemize}[leftmargin=1.5em]
    \item We introduce \textsc{SecProbe}, an \textbf{adaptive framework for evaluating coding agents' awareness of code vulnerabilities}. SecProbe follows an \emph{observe-then-act} loop: observed agent performance determines which security capability is probed next. Security evaluation thus becomes an iterative measurement process rather than a fixed task collection.
    
    \item We extend the adaptive assessment beyond fixed task banks through \textbf{on-demand task synthesis}. Guided by assessment needs, \textsc{SecProbe} constructs executable repository-scale tasks for target CWE classes and software contexts, together with exploit-based tests and reference patches, supplying new tasks when suitable ones are unavailable.
    
    \item We instantiate \textsc{SecProbe} with 353 highly challenging vulnerability-repair tasks spanning six programming languages and 151 distinct CWE types, and conduct a comprehensive empirical study of adaptive assessment, task quality, and the cybersecurity vulnerability awareness of nine frontier model backbones with two agent harnesses.
\end{itemize}

\section{Related Work}
\label{related}

\subsection{Code Security}

Executable security benchmarks assess coding agents across vulnerability discovery, exploitation, and repair. CVE-Bench evaluates real-world CVE repair in reproducible repositories \citep{wang-etal-2025-cve}, while CyberGym evaluates vulnerability reproduction through proof-of-concept generation \citep{wang2026cybergym}. ExploitGym extends vulnerability triggering to working exploits under deployed protections \citep{wang2026exploitgymaiagentsturn}, and CyberGym-E2E jointly evaluates discovery, proof-of-concept generation, repair, and post-patch functionality \citep{shi2026cybergymee}. Complementary work examines secure code generation and repair \citep{chen-etal-2026-securevibebench,mou-etal-2025-really} and broader attacks and defenses for tool-using agents \citep{fu2025rasevalcomprehensivebenchmarksecurity,zhang2025agent}. SEC-bench reproduces vulnerabilities in isolated repositories with evaluation harnesses and reference patches \citep{lee2025secbench}. CVE-Factory converts CVE metadata into executable environments \citep{luo2026cvefactory}, while AutoBaxBuilder synthesizes secure-code tasks and security tests from scratch \citep{arx2026autobaxbuilder}. These approaches expand the supply of executable security tasks. \textsc{SecProbe} connects task construction to \textbf{adaptive evaluation of coding-agent vulnerability awareness}: observed outcomes guide task selection and on-demand synthesis conditioned on target CWE classes and software contexts. Assessment needs thus direct both evaluation effort and task generation. Table~\ref{tab:benchmark_comparison} summarizes these comparisons.

\begin{table}[t]
\centering
\caption{Comparison of cybersecurity benchmarks and frameworks by task count, CWE coverage, and support for difficulty and cost control.}
\label{tab:benchmark_comparison}
\small
\setlength{\tabcolsep}{9pt}
\zebra
\begin{tabular}{@{}lrrcc@{}}
\headrow
\thd{Benchmark / Framework} & \thd{\# Tasks} & \thd{\# CWEs} & \thd{Difficulty control} & \thd{Cost control}\\
CVE-Bench~\citep{wang-etal-2025-cve} & 509 & 16 & \xmark & \xmark\\
SEC-bench~\citep{lee2025secbench} & 200 & 25 & \xmark & \xmark\\
CyberGym~\citep{wang2026cybergym} & 1,507 & 88 & \xmark & \xmark\\
CVE-Factory~\citep{luo2026cvefactory} & 190 & 74 & \xmark & \xmark\\
AutoBaxBuilder~\citep{arx2026autobaxbuilder} & 40 & 11 & \cmark & \xmark\\
\rowcolor{oursrow}
\textbf{\textsc{SecProbe} (Ours)} & 353 & \textbf{151} & \cmark & \cmark\\
\end{tabular}
\end{table}

\subsection{LLMs in Synthetic Data}

LLMs have demonstrated strong capabilities for producing synthetic data, with recent studies emphasizing the importance of generation quality, diversity, complexity, and reasoning-guided orchestration \citep{liu2024best,havrilla2024surveying,davidson2025orchestrating}. Building on earlier dataset-generation methods based on pretrained language models \citep{schick2021generating}, modern LLMs have enabled synthetic data creation across multilingual question answering \citep{riabi-etal-2021-synthetic}, instruction tuning \citep{xu2024magpie,zhang2025oasis,zhong2024synthet2c}, factual alignment \citep{wei2023simple}, and scientific domains \citep{huang2025chemorch}. Synthetic generation has also been used to expand data diversity while maintaining utility \citep{dai2025auggpt,chung-etal-2023-increasing,riaz2025metasynth}. More general frameworks such as DataGen unify dataset specification and generation \citep{huang2025datagen}, while RefineLab automatically optimizes dataset-refinement workflows to improve data quality \citep{luo2026refinelab}. System-message synthesis has also been shown to support broad and personalized alignment \citep{lee2024aligning}. Our setting extends synthetic data generation to executable repositories containing vulnerabilities and associated exploitation samples. \textsc{SecProbe} coordinates task construction and validation through specialized agents, using evaluation outcomes to guide subsequent task generation.

\subsection{Efficient Benchmarking}

Efficient benchmarking seeks reliable estimates of model performance from fewer evaluations. \citet{ye2023predictable} use matrix completion to predict unobserved model--task scores from a subset of evaluation records. \citet{perlitz2023efficient} study the trade-off between cost and reliability, showing that example subsampling can preserve model rankings when evaluated scenarios remain sufficiently diverse. \citet{vivek2023anchor} select representative instances by clustering model confidence profiles, estimating performance from a small set of anchor points. These approaches improve efficiency within an existing task collection. \textsc{SecProbe} combines adaptive task selection with on-demand synthesis, using IRT estimates to identify where the available tasks provide insufficient information. It then directs multi-agent generation toward those assessment needs, allowing both the evaluation schedule and the task pool to adapt to observed agent performance.

\section{SecProbe}
\label{sec:methodology}

\subsection{Preliminaries}
\label{sec:preliminaries}

We evaluate the vulnerability awareness of coding agents through repository-scale
tasks. Each task requires the agent to locate and repair vulnerabilities while
preserving the intended functionality.

Let $c\in\mathcal{C}$ denote a target CWE class, $s$ a software specification
describing the language, application context, and architecture, and $h$ a
difficulty-control vector (Section~\ref{sec:difficulty-control}). Given
$(c,s,h)$, \textsc{SecProbe} first builds an executable project $P$. It then
injects vulnerabilities of class $c$ into $P$, yielding $P^c=\mathcal{I}_c(P)$.
A task is represented as
\begin{equation}
    x=(s,P^c,T_{\mathrm{func}},T_{\mathrm{sec}},m),
\end{equation}
where $T_{\mathrm{func}}$ is a suite of unit tests covering the intended
functionality, $T_{\mathrm{sec}}$ is a suite of exploitation
tests, and $m$ records the task metadata. Throughout, a suite evaluates to one
when it passes and zero when it fails. Each task is further associated with a
ground-truth reference patch $\Delta^\star$. Both $T_{\mathrm{sec}}$ and
$\Delta^\star$ are hidden from the agent at test time. The tests in
$T_{\mathrm{sec}}$ are designed to trigger the injected vulnerabilities: a
failing suite ($T_{\mathrm{sec}}(P^c)=0$) means the exploit succeeded, and a
passing suite means it did not. A valid vulnerable sample must therefore behave
as the specification intends while remaining exploitable:
\begin{equation}
    T_{\mathrm{func}}(P^c)=1,
    \qquad
    T_{\mathrm{sec}}(P^c)=0 .
\end{equation}

The evaluated agent receives $s$ and $P^c$, with the exact vulnerability locations, security test samples, and the reference patch withheld. It must detect the vulnerabilities and produce a patch $\Delta$. Evaluation relies primarily on executing the security and unit tests against the patched repository $P^c\oplus\Delta$, where $\oplus$ denotes patch application. To satisfy the test-based repair criterion, the patch must meet
\begin{equation}
    T_{\mathrm{func}}(P^c\oplus\Delta)=1
    \quad\text{and}\quad
    T_{\mathrm{sec}}(P^c\oplus\Delta)=1.
\end{equation}
These conditions require preserving tested functionality and blocking the tested exploits. Supplementary review uses $\Delta^\star$ as a reference to assess whether the submitted changes address the underlying vulnerabilities.

Task synthesis defines a configurable distribution $x\sim\mathcal{D}_{\mathrm{gym}}(c,s,h)$ over executable evaluation tasks. The configuration controls vulnerability class, software context, and difficulty, allowing \textsc{SecProbe} to construct new tasks without requiring a pre-existing vulnerable repository.

\subsection{Overview}
\label{sec:cybergym-overview}

\begin{figure}[t]
  \centering
  \includegraphics[width=\textwidth]{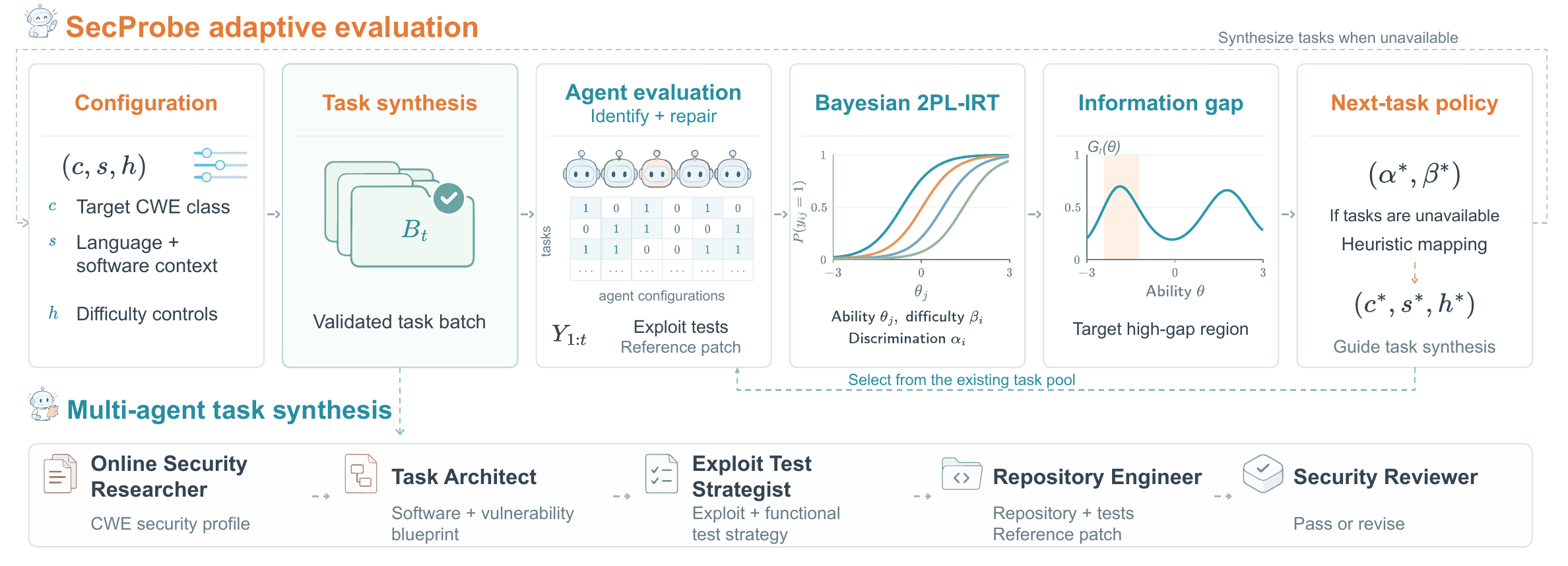}
  \caption{\textsc{SecProbe} couples adaptive evaluation with on-demand multi-agent task synthesis.}
  \label{fig:pipeline}
\end{figure}

As illustrated in Figure~\ref{fig:pipeline}, \textsc{SecProbe} connects adaptive evaluation with multi-agent task synthesis through two feedback loops. The outer loop determines which tasks to evaluate next. After each evaluation batch, an IRT model is fitted to accumulated agent outcomes to estimate agent ability, task difficulty, and discrimination. These estimates identify where additional evidence would be informative and guide selection from the available task pool. When suitable tasks are unavailable, the framework selects a generation configuration $(c,s,h)$ and invokes task synthesis. Outcomes on the selected and newly generated tasks update the model for the next round, making evaluation needs guide the expansion of the task pool (Section~\ref{sec:adaptive-synthesis}).

The inner loop constructs and reviews each new task through five specialized agents, the \emph{Online Security Researcher}, \emph{Task Architect}, \emph{Exploit Test Strategist}, \emph{Repository Engineer}, and \emph{Security Reviewer}. Together, they translate the target CWE class and software specification into an executable repository, reference patch, and evaluation tests. An orchestrator coordinates intermediate artifacts and routes issues identified during review back to the responsible agents for revision. Only tasks satisfying the release requirements enter the evaluation pool. Construction and testing take place in isolated synthetic environments. We next describe the specialist agents, difficulty controls, and adaptive evaluation procedure.

\subsection{Specialist Agents}
\label{sec:specialist-agents}

\textbf{Online Security Researcher.}
Given a target CWE class, this agent consults authoritative online sources to
understand how the vulnerability arises, how it is exploited in practice, and
what constitutes an effective mitigation.  It distills the findings into a
structured security profile $\phi_c=\mathcal{R}(c)$, where $\mathcal{R}$ denotes
the research process.  The profile $\phi_c$ grounds task synthesis in real
security behavior without reproducing attacks against live systems.

\textbf{Task Architect.}
This agent transforms the security profile $\phi_c$ and generation
specification $s$ into a repository-scale task blueprint
$b=\mathcal{A}(\phi_c,s)$.  It places the target vulnerability in a realistic
application and designs the surrounding software so that the security behavior
emerges through meaningful interactions across the project, rather than through
an isolated or easily identifiable code snippet.  The resulting blueprint $b$
serves as the shared contract for implementation and evaluation.

\textbf{Exploit Test Strategist.}
Using the blueprint and security profile, this agent develops a test strategy $q=\mathcal{S}(b,\phi_c)$. It specifies exploitation samples targeting the planned vulnerabilities and functional checks for legitimate use, guiding the construction of $T_{\mathrm{sec}}$ and $T_{\mathrm{func}}$. Together, these checks assess whether repairs block the attacks while retaining the application's intended behavior.

\textbf{Repository Engineer.}
Conditioned on $b$ and $q$, this agent constructs the executable instance
$x=\mathcal{E}(b,q)$.  It realizes the intended application and plants the
target vulnerabilities within plausible code paths while preserving the utility of the repo-level sample.  The agent then builds and exercises the project so that
$x=(s,P^c,T_{\mathrm{func}},T_{\mathrm{sec}},m)$ is reproducible and ready for
independent review.

\textbf{Security Reviewer.}
This agent independently audits $x$ and returns a decision $v=\mathcal{V}(x)\in\{\mathrm{pass},\mathrm{revise}\}$. It checks whether the intended behavior is preserved, the injected vulnerabilities are exploitable, and the evaluation recognizes correct repairs. A $\mathrm{revise}$ decision includes specific findings for the responsible agent. Review findings guide revision, while final admission to the \textsc{SecProbe} task pool is determined by the deterministic validation checks described in Section~\ref{app:validation-repair}.

\subsection{Validation and Release}
\label{app:validation-repair}

\paragraph{Acceptance thresholds}
An instance is admitted only if the reference implementation scores exactly $100$ under its hidden grader, security-bearing rubric categories carry at least half of the 100 points, and every declared security mutation applied to the reference scores at most $80$. The released vulnerable repository must pass the functionality tests and obtain a hidden-grader score in $[15,75]$.

\paragraph{Validation and repair loop}
The $\mathrm{pass}/\mathrm{revise}$ decision of the Security Reviewer is
advisory; final admission is determined by a deterministic validator that checks
artifact completeness, rejects symlinks, verifies public-API parity and scale
constraints, isolates $T_{\mathrm{sec}}$ from the released repository, and
requires the reference implementation to score $100$ deterministically.  The
released repository must be fully implemented, pass $T_{\mathrm{func}}$, satisfy
the score thresholds above, and document the required source-to-sink paths. The hidden grader must detect every declared mutant.  The exported task is then rebuilt and regraded in a
clean Linux container.  If validation fails, the orchestrator routes a
machine-readable failure report to a repair stage and reruns the complete
validator, for at most ten rounds.  Repairs may not weaken tests, modify the
reference implementation to accommodate stale mutants, pad the repository to
meet scale constraints, or alter the export or grading harness to induce a pass;
instances that remain invalid after the final round are discarded.

\subsection{Difficulty Control}
\label{sec:difficulty-control}

Each task evaluates two coupled capabilities. Given the specification $s$ and vulnerable repository $P^c$, the agent must first \emph{identify} an unknown set of vulnerabilities $Z_c=\{z_1,\ldots,z_K\}$. It must then \emph{repair} them by producing a patch $\Delta$ that addresses every member of $Z_c$ while preserving intended functionality. Vulnerability locations and exploitation paths are withheld, requiring the agent to reason across the repository to locate the vulnerabilities and determine how to patch.

To control the difficulty of this process, we condition generation on the
hyperparameter vector
\begin{equation}
    h=(h_{\mathrm{repo}},h_{\mathrm{arch}},
    h_{\mathrm{vuln}},h_{\mathrm{eval}}).
\end{equation}
The \emph{repository-scale} vector
$h_{\mathrm{repo}}=(n_{\mathrm{loc}},n_{\mathrm{file}},n_{\mathrm{func}})$
controls the numbers of source-code lines, implementation files, and functions.
The \emph{architectural} vector
$h_{\mathrm{arch}}=(n_{\mathrm{comp}},n_{\mathrm{dep}},n_{\mathrm{api}})$
controls software modules or services, dependencies between them, and exposed
types, functions, or methods. The \emph{vulnerability} vector
$h_{\mathrm{vuln}}=(n_{\mathrm{vuln}},n_{\mathrm{path}},n_{\mathrm{mut}})$
controls injected vulnerabilities, exploit paths per vulnerability, and
security mutations detected by the security tests. Finally, the \emph{evaluation} vector
$h_{\mathrm{eval}}=(n_{\mathrm{pub}},n_{\mathrm{hid}},n_{\mathrm{adv}})$
controls agent-visible functionality tests, hidden functional and security
tests reserved for evaluation, and exploit families with distinct mechanisms
or security conditions. Together, these controls define
$x\sim\mathcal{D}_{\mathrm{gym}}(c,s,h)$, allowing repository scale and security
complexity to vary within a CWE class. Precise definitions and bounds appear
in Appendix~\ref{app:hyperparameters}.

\subsection{Adaptive Evaluation}
\label{sec:adaptive-synthesis}

\textsc{SecProbe} evaluates coding-agent vulnerability awareness through successive rounds of task selection, synthesis, and evaluation. To initialize the assessment, it constructs a broad task batch $\mathcal{B}_1$ spanning vulnerability classes $c$, software specifications $s$, and difficulty configurations $h$. A pool $\mathcal{A}$ of coding-agent configurations attempts these tasks, yielding $y_{ij}=1$ when agent $j$ successfully repairs task $i$ and $y_{ij}=0$ otherwise. Responses from agents with different capabilities support joint estimation of agent ability and task characteristics. After round $t$, we fit a Bayesian two-parameter logistic IRT model to the accumulated responses $Y_{1:t}$,
\begin{equation}
    \Pr(y_{ij}=1)
    =\sigma\!\left(\alpha_i(\theta_j-\beta_i)\right),
\end{equation}
where $\sigma$ is the logistic function, $\theta_j$ denotes agent ability, $\beta_i$ task difficulty, and $\alpha_i>0$ task discrimination. The posterior $q_t=q(\boldsymbol{\theta},\boldsymbol{\alpha},\boldsymbol{\beta}\mid Y_{1:t})$ captures uncertainty in both agent and task parameters.

The next round targets ability regions where the current agent pool is concentrated but the available tasks provide little information. We quantify this mismatch using the information-gap score
\begin{equation}
    G_t(\theta)=\frac{w_t(\theta)}{\epsilon+\mathbb{E}_{q_t}\!\left[\sum_i\alpha_i^2p_i(\theta)(1-p_i(\theta))\right]},\qquad w_t(\theta)=\frac{1}{|\mathcal{A}|}\sum_{j\in\mathcal{A}}\Pr_{q_t}\!\left(|\theta_j-\theta|\leq\delta\right),
\end{equation}
where $p_i(\theta)=\sigma(\alpha_i(\theta-\beta_i))$, the sum over $i$ includes calibrated tasks, and $\epsilon>0$ ensures numerical stability. The weight $w_t(\theta)$ is the posterior expected fraction of agents within distance $\delta$ of $\theta$, while the denominator measures the information supplied by those tasks. A large $G_t(\theta)$ therefore indicates an ability region that is relevant to the agent pool but insufficiently assessed. We set the target difficulty to $\beta^\star=\arg\max_\theta G_t(\theta)$ and choose the target discrimination $\alpha^\star$ from an upper quantile of estimated discrimination values for previously generated tasks near $\beta^\star$. Together, these targets specify where additional evidence is needed and how strongly the next tasks should distinguish agents with nearby abilities.

The target $(\alpha^\star,\beta^\star)$ guides selection from the existing task pool. When suitable tasks are unavailable, \textsc{SecProbe} invokes multi-agent synthesis to construct new ones. To translate the assessment target into a generation configuration, we initialize heuristic mappings from the first batch,
\begin{equation}
    \hat{\alpha}_t(c,s,h),\qquad\hat{\beta}_t(c,s,h),
\end{equation}
which predict task discrimination and difficulty from the target CWE class, software specification, and difficulty controls. We select $(c^\star,s^\star,h^\star)$ whose predicted parameters are closest to $(\alpha^\star,\beta^\star)$ and use this configuration to guide synthesis. After evaluating each new batch, we update the IRT posterior and the mappings, allowing observed outcomes to refine subsequent task selection and construction. Batch sizes decrease as information gaps contract, and assessment stops when the normalized maximum information gap falls below the threshold specified in Appendix~\ref{app:hyperparameters}. Section~\ref{app:irt-validation} examines the sensitivity of agent rankings and task selection to the IRT specification.

\section{Experiments}
\label{experiments}

\subsection{Experimental Setup}
\label{sec:experimental-setup}
\label{app:evaluation-protocol}

We implement \textsc{SecProbe} with the OpenAI Agents SDK, using GPT-5.6-Luna with medium reasoning effort for all specialist agents. Tasks span six programming languages (Python, C++, Go, Java, TypeScript, and Rust) and three software levels comprising library components, single-process applications, and multi-service applications. Each task is specified by a CWE class $c$, software specification $s$, and difficulty-control vector $h$. As one application of the framework, we conduct adaptive evaluation starting from 150 tasks and reduce subsequent batch sizes as information gaps contract, producing a pool of 353 accepted tasks. Appendix~\ref{app:implementation} provides complete implementation details.

\paragraph{Models}
We select nine recent models from leading frontier families as agent backbones: GPT-5.6 Sol~\citep{openai2026gpt56}, GPT-5.4-mini~\citep{openai2026gpt54mini}, GLM-5.3~\citep{zai2026glm53}, Kimi K3~\citep{kimiteam2026kimik3openfrontier}, DeepSeek-V4 Pro~\citep{deepseekai2026deepseekv4highlyefficientmilliontoken}, Qwen3.7-Flash~\citep{qwen2026qwen37}, Opus 4.8~\citep{anthropic2026claudeopus48}, MiniMax-M3~\citep{lai2026minimaxsparseattention}, and Mistral Small
4~\citep{mistralai2026mistralsmall4}. The resulting set covers eight providers and spans proprietary and open-weight models, as well as flagship and efficiency-oriented variants.

\paragraph{Agent evaluation}
We evaluate each backbone model with both \textsc{Mini-SWE-Agent}~\citep{yang2024sweagent} and the
\textsc{Terminus-2}~\citep{merrill2026terminalbench} harness.  Each run takes place in an isolated Linux
environment with a common task image and resource budget. Evaluating both harnesses lets us separate
model capability from the effect of the surrounding agent scaffold. For a given model--harness configuration, the agent
receives the same solver-facing task specification and vulnerable repository.
The secure reference implementation, hidden tests, vulnerability manifest, and
grader are not exposed during the run.  The agent may inspect and modify the
repository, execute commands, and run the public tests within \texttt{/app}.
After the agent terminates, its final workspace is evaluated once by the hidden
grader; no additional repair is allowed after grading begins. Reference patches additionally support review of whether the submitted changes address the injected vulnerabilities.

\paragraph{Execution environment}
Each run is executed in a fresh Linux container built from the task image.  We
allocate 2 CPU cores, 4,096\,MB of memory, and 10,240\,MB of storage per run.
Image construction is limited to 1,200\,s, agent execution to 7,200\,s, and
hidden verification to 600\,s.  Both harnesses use the same task image,
filesystem layout, resource limits, and grader.  The container is discarded
after evaluation, preventing files or processes from carrying over between
runs.

\paragraph{Scoring \& Metrics}
We report two metrics. \textbf{(1)} The \emph{pass rate} is the percentage of agent runs that obtain the full grader score of 100, indicating that the submitted patch satisfies all functional and security requirements; every partial score is treated as a failure. \textbf{(2)} The \emph{mean normalized grader score} averages the grader scores across task attempts for each model--harness configuration and captures partial progress when a task is not completely solved.

Each task grader returns a score $g_r\in[0,100]$ for run $r$, with rubric weights
shared by the vulnerable seed, secure reference, and agent submission.  For a
model--harness configuration with run set $\mathcal{R}$, we compute
\begin{equation}
    \operatorname{PassRate}
    = \frac{100}{|\mathcal{R}|}\sum_{r\in\mathcal{R}}
      \mathbf{1}\{g_r=100\},
    \qquad
    \operatorname{MeanScore}
    = \frac{1}{|\mathcal{R}|}\sum_{r\in\mathcal{R}} g_r.
\end{equation}
Only a score of $100$ is counted as a successful repair; partial scores
contribute to the mean score but not to the pass rate.  This strict criterion
requires the patch to satisfy every functional and security category rather
than only the most heavily weighted requirements.

\subsection{Quality Analysis of Generated Tasks}

\begin{figure}[t]
  \centering
  \includegraphics[width=\textwidth]{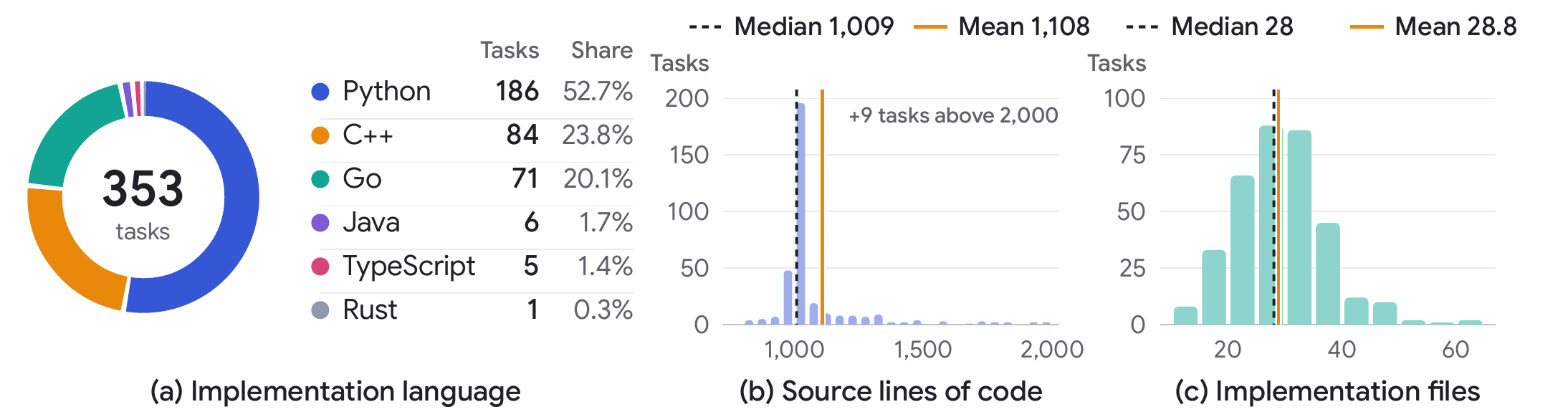}
  \caption{Coding characteristics of the 353-task pool, showing (a) implementation languages, (b) source lines of code, and (c) implementation-file counts.}
  \label{fig:coding-statistics}
\end{figure}

We examine whether the generated tasks provide suitable environments for evaluating vulnerability identification and repair through task statistics and expert review.

\begin{figure}[t]
  \centering
  \includegraphics[width=\textwidth]{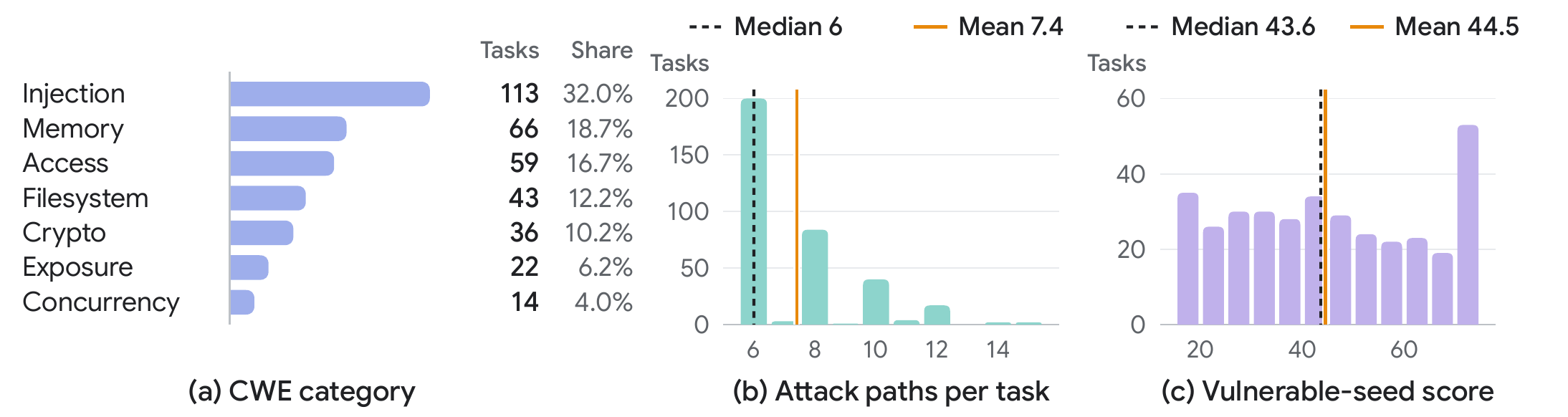}
  \caption{Security characteristics of the 353-task pool, showing (a) broad security categories, (b) distinct attack paths per task, and (c) grader scores before repair.}
  \label{fig:security-statistics}
  
\end{figure}

\paragraph{Task statistics}
Figures~\ref{fig:coding-statistics} and~\ref{fig:security-statistics} summarize the coding and security characteristics of the 353 generated tasks from the evaluation task pool, spanning six programming languages and 151 CWE types across seven broad security categories. Repositories contain a median of 1,009 source lines and 28 implementation files, while tasks expose a median of 6 distinct attack paths, requiring diagnosis and repair across multiple files and security requirements. Vulnerable seeds obtain a mean grader score of 44.5, leaving substantial room for repair before reaching the full score of 100. These characteristics suggest a challenging evaluation setting for coding agents, demanding both repository-level reasoning and comprehensive security repair.

\begin{figure}[t]
  \centering
  \includegraphics[width=\textwidth]{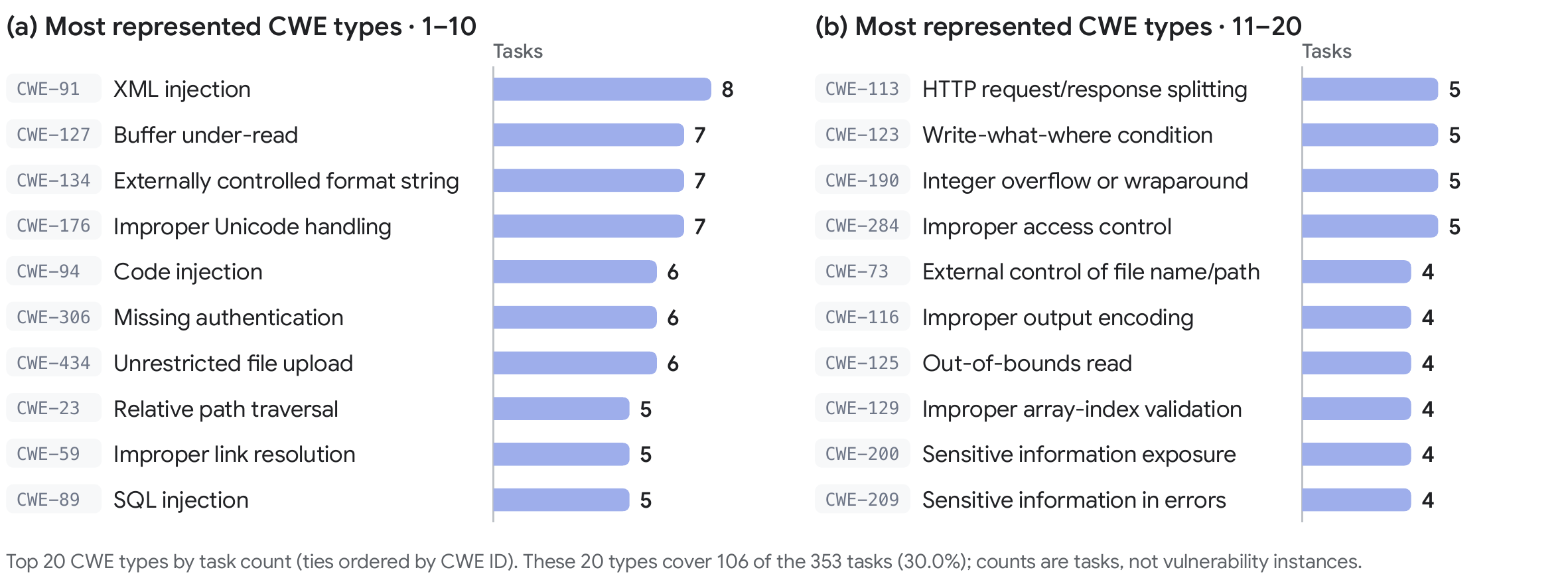}
  \caption{Detailed CWE coverage.}
  \label{fig:detailed-cwe-coverage}
\end{figure}

\paragraph{Detailed CWE coverage}
Figure~\ref{fig:detailed-cwe-coverage} presents the 20 most represented
individual CWE types, ranked by task count. Each task contributes once to its
configured CWE type, regardless of how many vulnerability instances it
contains. These 20 types account for 106 of the 353 tasks (30.0\%). XML
injection (CWE-91) is the most represented type, with eight tasks (2.3\%),
followed by buffer under-read (CWE-127), externally controlled format strings
(CWE-134), and improper Unicode handling (CWE-176), with seven tasks each.
The ranking also includes access-control, path-traversal, and
information-exposure weaknesses, revealing variation within the broad
security categories in Figure~\ref{fig:security-statistics}.

\begin{figure}[t]
  \centering
  \includegraphics[width=\textwidth]{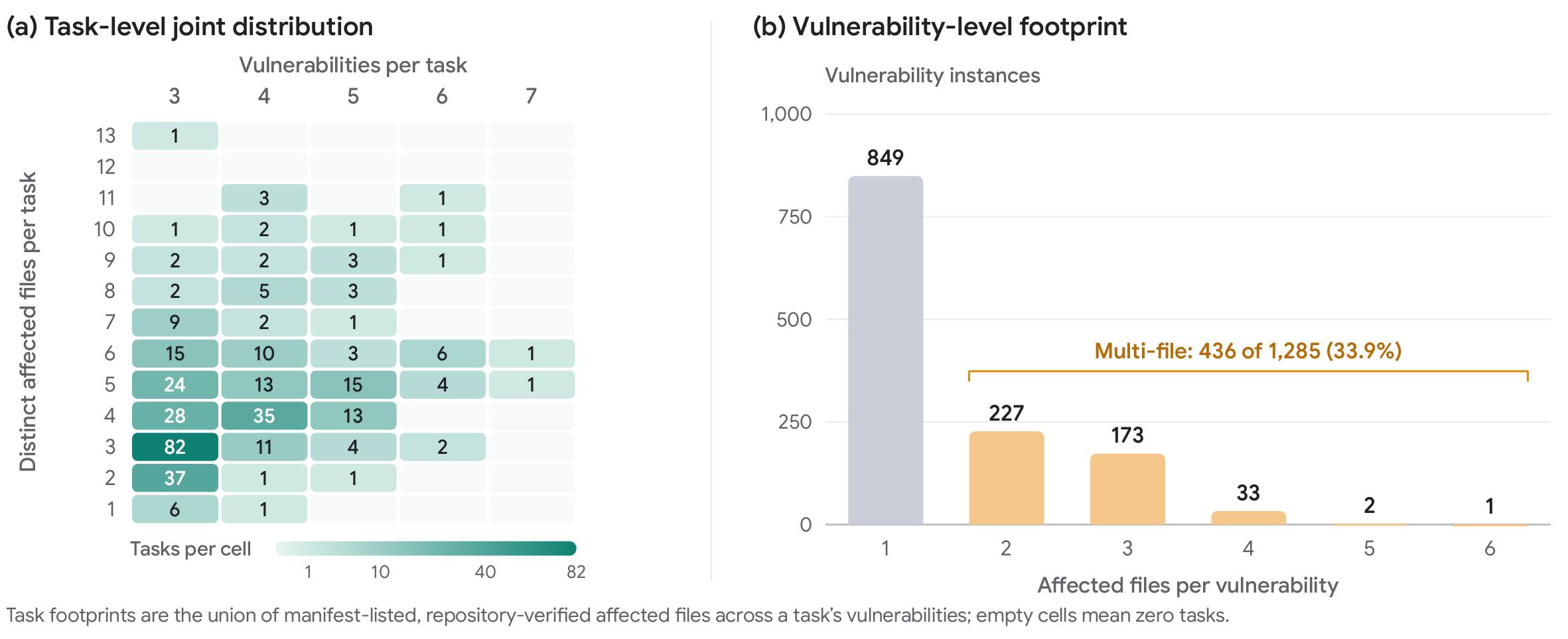}
  \caption{Cross-file security complexity.}
  \label{fig:cross-file-security-complexity}
\end{figure}

\paragraph{Cross-file security complexity}
We count the distinct manifest-listed files associated with each vulnerability
after resolving their paths against the repository. A task's affected-file
footprint is the union across its vulnerabilities.
Figure~\ref{fig:cross-file-security-complexity}(a) relates this footprint to
vulnerability count. Across 353 tasks, the median footprint is four files
(range: 1--13), and 346 tasks (98.0\%) affect multiple files. The most frequent
combination is three vulnerabilities and three affected files, with 82 tasks
(23.2\%). Tasks with three vulnerabilities nevertheless span 1--13 affected
files, so vulnerability count alone does not capture the breadth of affected
code. At the individual-vulnerability level, 436 of 1,285 instances (33.9\%)
list multiple files (Figure~\ref{fig:cross-file-security-complexity}(b)).
Task footprints may combine independent single-file vulnerabilities; they
describe documented affected-file coverage, not minimum patch sizes or runtime
execution paths.

\begin{figure}[t]
  \centering
  \includegraphics[width=\textwidth]{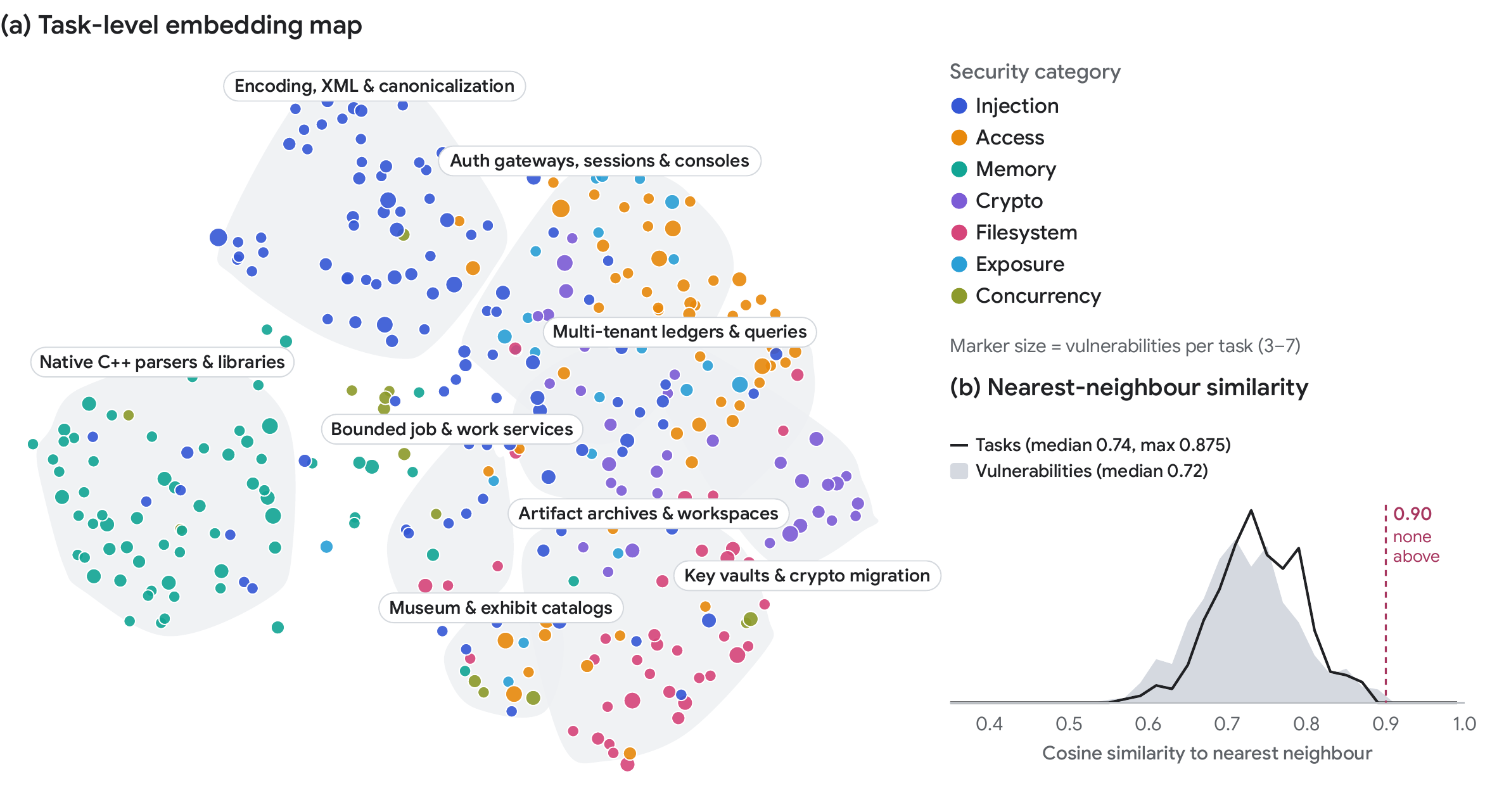}
  \caption{Semantic map of the 353-task pool. (a) Task specifications embedded with Qwen3-Embedding-8B and projected with UMAP. Colour gives the security category of each task's CWE, marker size its number of vulnerabilities, and shaded regions the eight application themes. (b) Cosine similarity of each task and each vulnerability to its nearest neighbour.}
  \label{fig:embedding-task-map}
\end{figure}

\paragraph{Semantic diversity}
To test whether synthesis yields distinct tasks rather than variants of a few templates, we embed each task specification (its title and application description, without the scale constraints shared by all tasks) and each of the 1,285 vulnerability descriptions (title and rationale) with Qwen3-Embedding-8B~\citep{zhang2025qwen3embedding}, and project both sets with UMAP~\citep{mcinnes2018umap} under cosine distance. CWEs are grouped into seven broad security categories. Figure~\ref{fig:embedding-task-map}(a) shows that the pool separates into eight application themes, obtained by Ward clustering~\citep{ward1963hierarchical} of an eight-dimensional UMAP projection; on average, 85\% of a task's five nearest neighbours in the map share its theme. Themes align with security categories where the software determines the weakness: 83\% of the native C++ library tasks target memory safety, and 92\% of the encoding and XML tasks target injection. At the same time, only 11.6\% of a task's ten nearest neighbours in the embedding space share its CWE, compared with 59.1\% that share its broad category and 19.4\% expected by chance. Tasks with the same CWE are thus placed in different applications, while related weaknesses remain locally coherent. No pair of tasks exceeds a cosine similarity of 0.90: the median nearest-neighbour similarity is 0.74 and the maximum is 0.875 (Figure~\ref{fig:embedding-task-map}(b)), so the pool contains no near-duplicate specifications.

\begin{figure}[t]
  \centering
  \includegraphics[width=\textwidth]{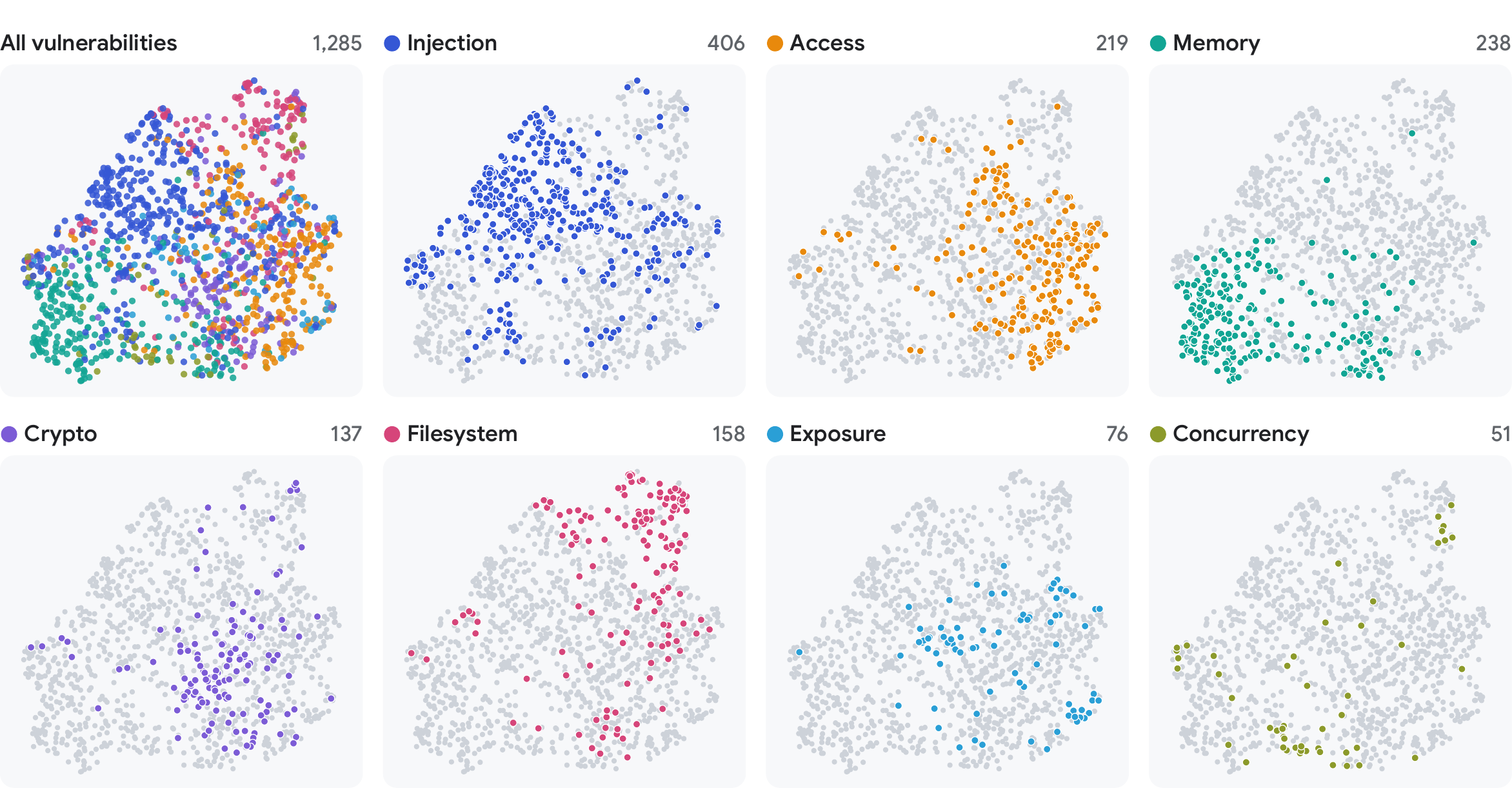}
  \caption{Vulnerability-level map of the 1,285 vulnerability descriptions. The first panel colours every vulnerability by security category; each remaining panel highlights one category against the others.}
  \label{fig:embedding-vuln-facets}
\end{figure}

\paragraph{Vulnerability-level diversity}
Figure~\ref{fig:embedding-vuln-facets} maps the 1,285 vulnerability descriptions. Each category occupies a core region but extends into others. Memory-safety vulnerabilities are the most concentrated, with 65\% of their ten nearest neighbours in the same category, whereas exposure and concurrency vulnerabilities are the most dispersed, at 26\% and 23\%. Vulnerabilities in the same task are no more similar to one another (mean cosine similarity 0.506) than to vulnerabilities of the same CWE in other tasks (0.503), so the three to seven vulnerabilities in a task are distinct flaws rather than restatements of one. These findings do not depend on the embedding model: with OpenAI text-embedding-3-large, pairwise task similarities correlate with those above at Spearman $\rho=0.75$, and no task pair exceeds 0.90.

\paragraph{Human evaluation}
To assess task quality beyond automated checks, experts evaluate security validity, attack plausibility and diversity, reference-solution correctness, repository realism, cross-file complexity, and grader alignment with functional and security requirements. Mean ratings exceed 4.0 out of 5 in every dimension, with an overall rating of 4.40 and 92\% of evaluated tasks accepted for benchmark inclusion. These assessments suggest that the generated repositories resemble real-world software in structure, and that their injected vulnerabilities represent valid security weaknesses with plausible attack paths. Table~\ref{tab:human-quality} reports the full results.

\label{app:human-evaluation}
\paragraph{Human evaluation protocol}
For each selected task, experts assess whether the injected vulnerabilities match the target CWE and can be exploited (\emph{security validity}); whether attack scenarios are plausible and non-equivalent (\emph{attack plausibility and diversity}); whether the reference repair blocks all documented attack paths without regressions (\emph{reference correctness}); whether the repository forms a coherent software project (\emph{repository realism}); whether repair requires reasoning across files or components (\emph{cross-file complexity}); and whether the tests and rubric accept complete alternative repairs while rejecting incomplete ones (\emph{grader alignment}).

\paragraph{Ratings and aggregation}
Experts score each dimension from 1 (poor) to 5 (excellent).  We report the mean
and standard deviation of these ratings, the proportion of ratings at least 4,
the proportion of positive acceptance judgments, and agreement across expert
judgments.  As shown in Table~\ref{tab:human-quality}, every criterion averages
above 4.0; the overall mean is $4.40$, with 92\% acceptance and an agreement
score of $0.80$.

\begin{table}[t]
\centering
\caption{Human evaluation of generated cybersecurity coding tasks. Experts score each dimension from 1 (poor) to 5 (excellent).}
\label{tab:human-quality}
\small
\setlength{\tabcolsep}{6pt}
\zebra
\begin{tabular}{@{}llcccc@{}}
\headrow
 & & \multicolumn{3}{c}{\thd{Expert ratings}} & \\
\headrow
\thd{Dimension} & \thd{Evaluation focus} & \thd{Mean $\pm$ Std.} & \thd{$\geq 4$} & \thd{Accept} & \thd{Agreement}\\
Security & CWE validity and exploit impact & 4.60 $\pm$ 0.50 & 92\% & 94\% & 0.82 \\
Attacks & Plausible and diverse attack cases & 4.20 $\pm$ 0.70 & 85\% & 87\% & 0.74 \\
Reference & Complete fix without regressions & \textbf{4.70 $\pm$ 0.40} & \textbf{94\%} & \textbf{96\%} & \textbf{0.86} \\
Realism & Coherent real-world repository & 4.10 $\pm$ 0.70 & 82\% & 84\% & 0.72 \\
Complexity & Cross-file diagnosis and repair & 4.40 $\pm$ 0.60 & 89\% & 91\% & 0.79 \\
Grader & Functional--security score alignment & 4.50 $\pm$ 0.50 & 91\% & 93\% & 0.83 \\
\rowcolor{oursrow}
\textbf{Overall} & \textbf{Benchmark inclusion suitability} & \textbf{4.40 $\pm$ 0.50} & \textbf{89\%} & \textbf{92\%} & \textbf{0.80} \\
\end{tabular}
\end{table}

\subsection{Evaluation of Coding-Agent Vulnerability Awareness}

We use the 353-task pool to assess how effectively frontier coding agents identify and repair vulnerabilities under a common execution and scoring protocol.

\begin{figure}[t]
  \centering
  \includegraphics[width=\textwidth]{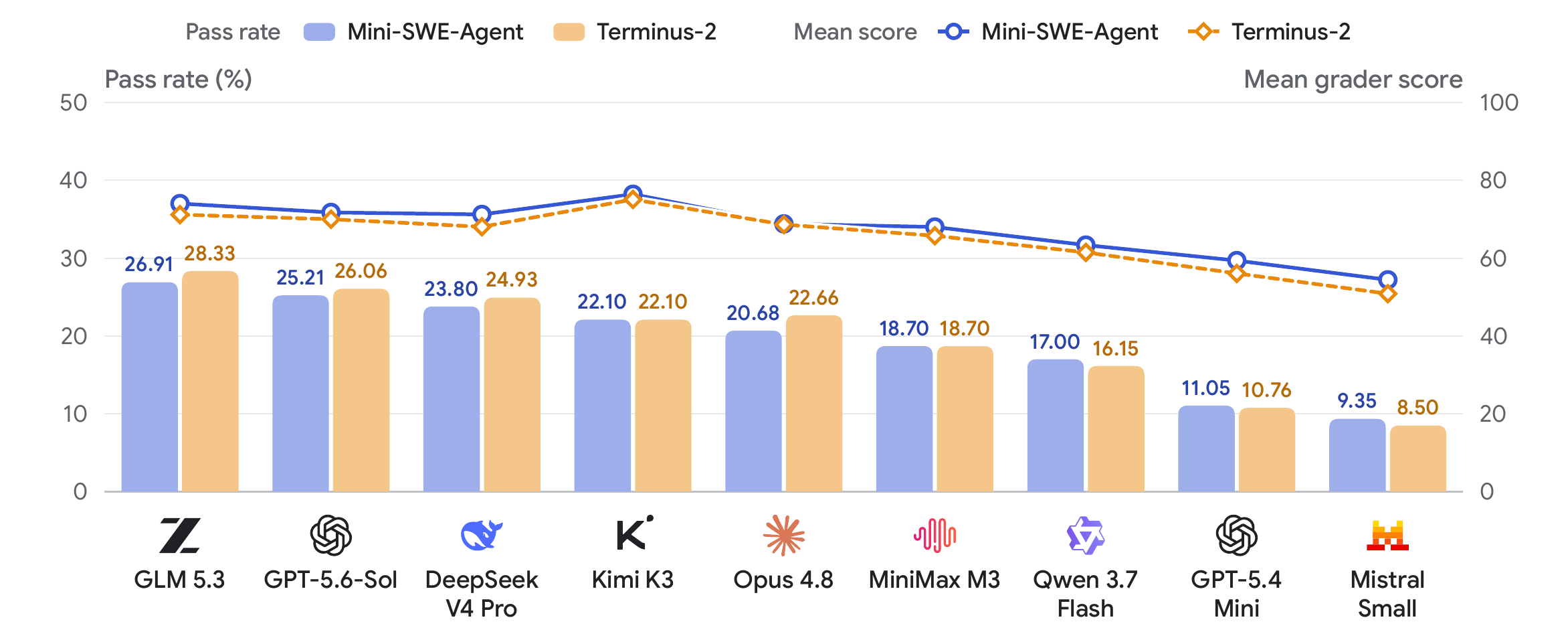}
  \caption{Agent performance on the 353-task pool with \textsc{Mini-SWE-Agent} and \textsc{Terminus-2}. Bars show strict pass rates; lines show mean normalized grader scores.}
  \label{fig:frontier-agent-results}
\end{figure}

\paragraph{Frontier agents remain far from reliable security repair}
Figure~\ref{fig:frontier-agent-results} shows that no evaluated configuration achieves a strict pass rate above 30\%. GLM-5.3 performs best under both harnesses, passing 26.91\% of runs with \textsc{Mini-SWE-Agent} and 28.33\% with \textsc{Terminus-2}. GPT-5.6-Sol ranks second, reaching 25.21\% with \textsc{Mini-SWE-Agent} and 26.06\% with \textsc{Terminus-2}. DeepSeek-V4 Pro follows with pass rates of 23.80\% and 24.93\% under the two harnesses. At the other end, the efficiency-oriented GPT-5.4-mini and Mistral Small remain below 12\% under both harnesses. The nearly 20-point spread between the strongest and weakest configurations shows that the benchmark meaningfully differentiates current model capabilities, while the low absolute pass rates leave substantial room for future progress.

\paragraph{Fine-grained repair performance}
To examine agent performance beyond the binary pass criterion, we also report the mean normalized grader score. As shown previously in Figure~\ref{fig:security-statistics}, the vulnerable repositories obtain a mean score of 44.5 before repair. After agent intervention, the mean scores range from 50.87 to 76.43 across model--harness configurations, showing that agents often make measurable progress even when they do not achieve a complete repair. Nevertheless, the weakest configuration improves over the vulnerable seed by only 6.37 points, and even the strongest mean score remains 23.57 points below a fully correct repair. Thus, frontier agents recover some functional and security behavior but still leave a substantial gap to complete
verification. Kimi K3 obtains the highest mean score under both harnesses, reaching 76.43 with \textsc{Mini-SWE-Agent} and 75.07 with \textsc{Terminus-2}, whereas GLM-5.3 achieves the highest pass rate.

\paragraph{Harness choice affects absolute performance but not the broad ranking}
The two harnesses produce broadly consistent model orderings, but their absolute behavior differs. \textsc{Terminus-2} achieves a higher pass rate for several models, including GLM-5.3, GPT-5.6-Sol, and DeepSeek-V4 Pro, while \textsc{Mini-SWE-Agent} obtains a higher mean grader score across the evaluated backbones. Thus, a harness can change the likelihood of reaching an exact solution without uniformly improving partial coverage of the grader requirements.

\subsection{Patch-Level Review}
\label{app:patch-level-evaluation}

\paragraph{Scope and comparison}
Across all 353 tasks, we supplement executable tests with reference-based assessment of agent patches using GPT-5.6-Luna. For each submitted repair $\Delta$, we obtain two independent judgments in separate model calls. Each call receives the same task requirements, relevant repository context, submitted changes, reference patch $\Delta^\star$, and test results, without access to the other judgment. The reference establishes the intended security requirements; alternative implementations can receive favorable judgments when they satisfy those requirements. The comparison concerns repair semantics rather than textual similarity.

\paragraph{Judging criteria}
Both judgments use the same rubric to assess whether the patch addresses the vulnerability's root cause, covers the affected functions and execution paths, and preserves required application behavior. Each returns a repair-correctness rating on a 0--100 scale and a concise rationale grounded in specific code changes or remaining omissions. A rating of 0 denotes no effective repair, while 100 denotes a complete repair satisfying the task requirements. This review complements exploit-based testing by examining whether a patch enforces the underlying security requirement beyond the particular inputs exercised by the tests.

\paragraph{Disagreement and expert review}
Let $r_1,r_2\in[0,100]$ denote the two ratings. A submission is referred to a security expert when $|r_1-r_2|>20$, corresponding to a disagreement threshold of $\tau=20$ points. The expert reviews the submission, reference patch, test evidence, and both rationales to adjudicate the disagreement. The two ratings and any expert decision are retained with their supporting evidence. Patch-level judgments provide supplementary evidence of repair correctness; the reported pass rates and mean evaluation scores follow the executable scoring protocol in Section~\ref{sec:experimental-setup}.

\paragraph{Review outcomes}
Table~\ref{tab:patch-level-review} presents outcomes for one submitted patch per task. We report the proportion of judge ratings within the disagreement threshold and the fraction requiring expert review. We separately report expert judgments of incomplete repair, including cases that pass the executable tests. The latter distinguishes test success from reference-based assessment of whether the underlying security requirements are fully addressed.

\begin{table}[t]
\centering
\caption{Patch-level review outcomes.}
\label{tab:patch-level-review}
\small
\setlength{\tabcolsep}{9pt}
\zebra
\begin{tabular}{@{}lrr@{}}
\headrow
\thd{Review outcome} & \thd{Count / Total} & \thd{Share (\%)}\\
Patches assessed by both judges & 353 / 353 & 100.0 \\
Judge ratings within tolerance ($|r_1-r_2|\leq20$) & 312 / 353 & 88.4 \\
Referred for expert review ($|r_1-r_2|>20$) & 41 / 353 & 11.6 \\
Expert-reviewed patches judged incomplete & 29 / 41 & 70.7 \\
Test-passing patches judged incomplete by an expert & 8 / 100 & 8.0 \\
\end{tabular}
\end{table}

\subsection{Efficiency of Adaptive Evaluation}
\label{sec:efficiency}

\begin{figure}[t]
    \centering
    \includegraphics[width=\linewidth]{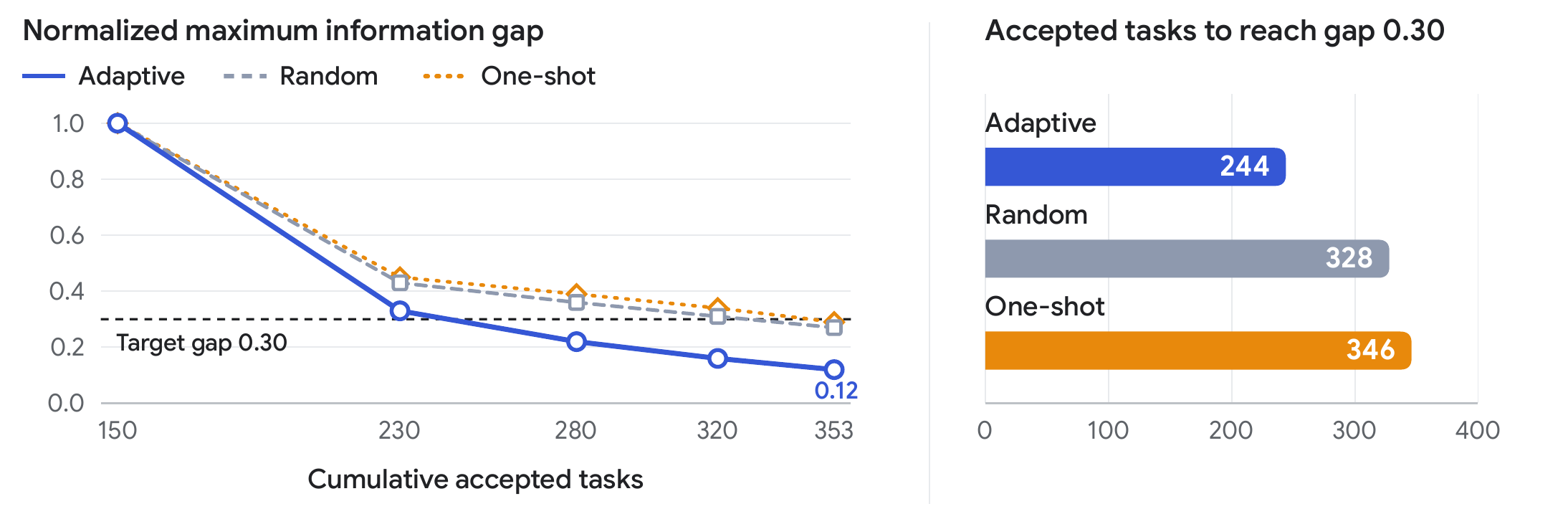}
    \caption{Adaptive evaluation versus random and one-shot baselines. Left shows the normalized maximum information gap as tasks accumulate; right shows the accepted tasks needed to reach a gap of 0.30.}
    \label{fig:adaptive-synthesis-effectiveness}
\end{figure}

\textbf{Does adaptive evaluation reduce the number of tasks needed?}
We measure efficiency using the information-gap function $G_t(\theta)$
defined in Section~\ref{sec:adaptive-synthesis}. We report the normalized
maximum gap
$\widetilde{G}_t=
\frac{\max_{\theta\in\Theta}G_t(\theta)}
{\max_{\theta\in\Theta}G_1(\theta)}$,
where $\Theta$ is a fixed grid spanning the ability range estimated from
the shared initial batch. Lower values indicate a smaller worst-case
information gap. Figure~\ref{fig:adaptive-synthesis-effectiveness} compares
our IRT-guided policy with random synthesis and one-shot synthesis. All strategies start
with the same 150 tasks and use identical validation procedures.
Adaptive synthesis reaches the target $\widetilde{G}_t\leq0.30$ after
244 accepted tasks, compared with 328 for random and 346 for one-shot
synthesis. These correspond to reductions of 25.6\% and 29.5\%, respectively. Agent feedback
thus helps target under-measured ability regions with fewer accepted tasks.

\paragraph{Held-out agents}
\label{app:held-out-agents}
We examine whether the evaluation efficiency of \textsc{SecProbe} transfers to models that do not participate in task construction or calibration. We partition the nine backbones into six construction models and three held-out models, keeping both harnesses of each backbone in the same partition. Only the construction models provide feedback for adaptive synthesis and item calibration. The resulting 353-task pool and item-parameter posterior are fixed before evaluating the six held-out model--harness configurations. Their responses update only their own ability estimates, without revising item parameters or generating further tasks.

\begin{table}[t]
\centering
\caption{Ability-estimation efficiency on held-out agents. RMSE is relative to full-pool ability estimates.}
\label{tab:held-out-agent-efficiency}
\small
\setlength{\tabcolsep}{9pt}
\zebra
\begin{tabular}{@{}lccc@{}}
\headrow
 & \multicolumn{2}{c}{\thd{Ability RMSE $\downarrow$}} & \\
\headrow
\thd{Policy} & \thd{100 tasks} & \thd{200 tasks} & \thd{Tasks to reach RMSE $\leq 0.15$ $\downarrow$}\\
\rowcolor{oursrow}
Adaptive & 0.180 & 0.110 & 140 \\
Random & 0.240 & 0.140 & 181 \\
Fixed order & 0.260 & 0.145 & 194 \\
\end{tabular}
\end{table}

\paragraph{Assessment protocol}
All policies start with the same 20 tasks and select without replacement from the same pool. Adaptive selection maximizes expected Fisher information under the current ability posterior, accounting for uncertainty in the fixed item calibration. Baselines use uniform random selection or a fixed ordering by expected information under the initial ability prior. Across 20 shared initialization seeds, we compare ability estimates with posterior means from all 353 tasks, used solely as assessment references. Task selection uses only responses observed so far.

\paragraph{Efficiency on unseen models}
Table~\ref{tab:held-out-agent-efficiency} reports ability RMSE across held-out configurations and seeds at budgets of 100 and 200 tasks, and the smallest budget beyond which RMSE stays at or below 0.15. Errors use the calibration ability scale, and all budgets include the initial 20 tasks. In the comparison, adaptive selection reaches the target with 140 tasks, versus 181 for random selection and 194 for the fixed ordering, with lower error at both common budgets.

\paragraph{Robustness to the IRT specification}
\label{app:irt-validation}
We examine whether the choice of IRT model changes agent rankings or adaptive task selection. We compare the unidimensional 2PL model used by \textsc{SecProbe} with Rasch (1PL), which fixes item discrimination, and a hierarchical 2PL model that allows agent-specific strengths across languages and security categories,
\begin{equation}
    p_{ij}=\sigma\!\left[\alpha_i\left(\theta_j+u_{j,\ell(i)}+v_{j,g(i)}-\beta_i\right)\right],
\end{equation}
where $\ell(i)$ and $g(i)$ denote task $i$'s language and security category. Centered, partially pooled effects $u$ and $v$ capture deviations from overall ability. All models use the same observed responses, candidate pool, and selection policy. We measure agreement with 2PL using Spearman correlation of posterior-mean overall ability rankings and the fraction of shared tasks among the next 20 selections.

\begin{table}[t]
\centering
\caption{Robustness of agent rankings and task selection to the IRT specification.}
\label{tab:irt-validation}
\small
\setlength{\tabcolsep}{9pt}
\zebra
\begin{tabular}{@{}lcc@{}}
\headrow
\thd{Model} & \thd{Ability rank Spearman $\rho$} & \thd{Top-20 selection overlap (\%)}\\
Rasch (1PL) & 0.96 & 80 \\
2PL & \textcolor{muted}{Reference} & \textcolor{muted}{Reference} \\
Hierarchical 2PL & 0.98 & 90 \\
\end{tabular}
\end{table}

As shown in Table~\ref{tab:irt-validation}, Rasch and hierarchical 2PL yield rank correlations of 0.96 and 0.98 with 2PL, respectively, and retain 16 and 18 of its 20 selected tasks. This agreement indicates that agent rankings and task selection are relatively stable across simpler and richer IRT specifications.

\subsection{Validating Difficulty Control}
\label{sec:difficulty-validation}

\textbf{Can \textsc{SecProbe} reliably control task difficulty?}
We evaluate difficulty control using 30 matched task triplets.  Within each
triplet, the CWE class and software specification are fixed, while the
difficulty-control vector is set to Easy, Medium, or Hard.  Each of the 90 task
variants is attempted five times by GPT-5.6-Sol with
\textsc{Mini-SWE-Agent}, producing 450 runs under the same execution and scoring
protocol.  We compare pass rate and mean grader score across the three levels
and record whether each triplet follows the intended ordering.  For the
one-factor interventions, we vary repository, vulnerability, or evaluation
complexity while holding the remaining controls fixed, allowing the effect of
each control family to be measured separately.

As shown in Table~\ref{tab:difficulty_control}, the pass rate decreases from 44.0\% to 13.3\% and the mean grader score from 78.6 to 55.8 from Easy to Hard, while 73.3\% of matched triplets follow the complete requested ordering. One-factor interventions further show that repository, vulnerability, and evaluation controls each increase empirical difficulty, with vulnerability complexity producing the largest effect.

\begin{table}[t]
\centering
\caption{Difficulty-control validation.}
\label{tab:difficulty_control}
\small
\setlength{\tabcolsep}{12pt}
\begin{tabular}{@{}lccc@{}}
\headrow
\thd{Setting} & \thd{Pass rate (\%) $\downarrow$} & \thd{Mean score $\downarrow$} & \thd{Ordered (\%) $\uparrow$}\\
\tgroup{4}{Overall difficulty levels}\\
Easy & $44.0_{\pm 3.2}$ & $78.6_{\pm 2.0}$ & -- \\
\rowcolor{tblzebra}
Medium & $27.3_{\pm 2.8}$ & $67.3_{\pm 2.2}$ & 83.3 \\
Hard & $13.3_{\pm 2.1}$ & $55.8_{\pm 2.4}$ & 80.0 \\
\tgroup{4}{Individual complexity effects}\\
Repository complexity & $-8.0_{\pm 3.1}$ & $-5.4_{\pm 2.2}$ & 73.3 \\
\rowcolor{tblzebra}
Vulnerability complexity & $-13.3_{\pm 3.8}$ & $-9.1_{\pm 2.7}$ & 86.7 \\
Evaluation complexity & $-10.0_{\pm 3.4}$ & $-7.3_{\pm 2.4}$ & 80.0 \\
\end{tabular}
\end{table}

\subsection{Failure Analysis}
\label{sec:failure-analysis}

\begin{figure}[t]
  \centering
  \includegraphics[width=\textwidth]{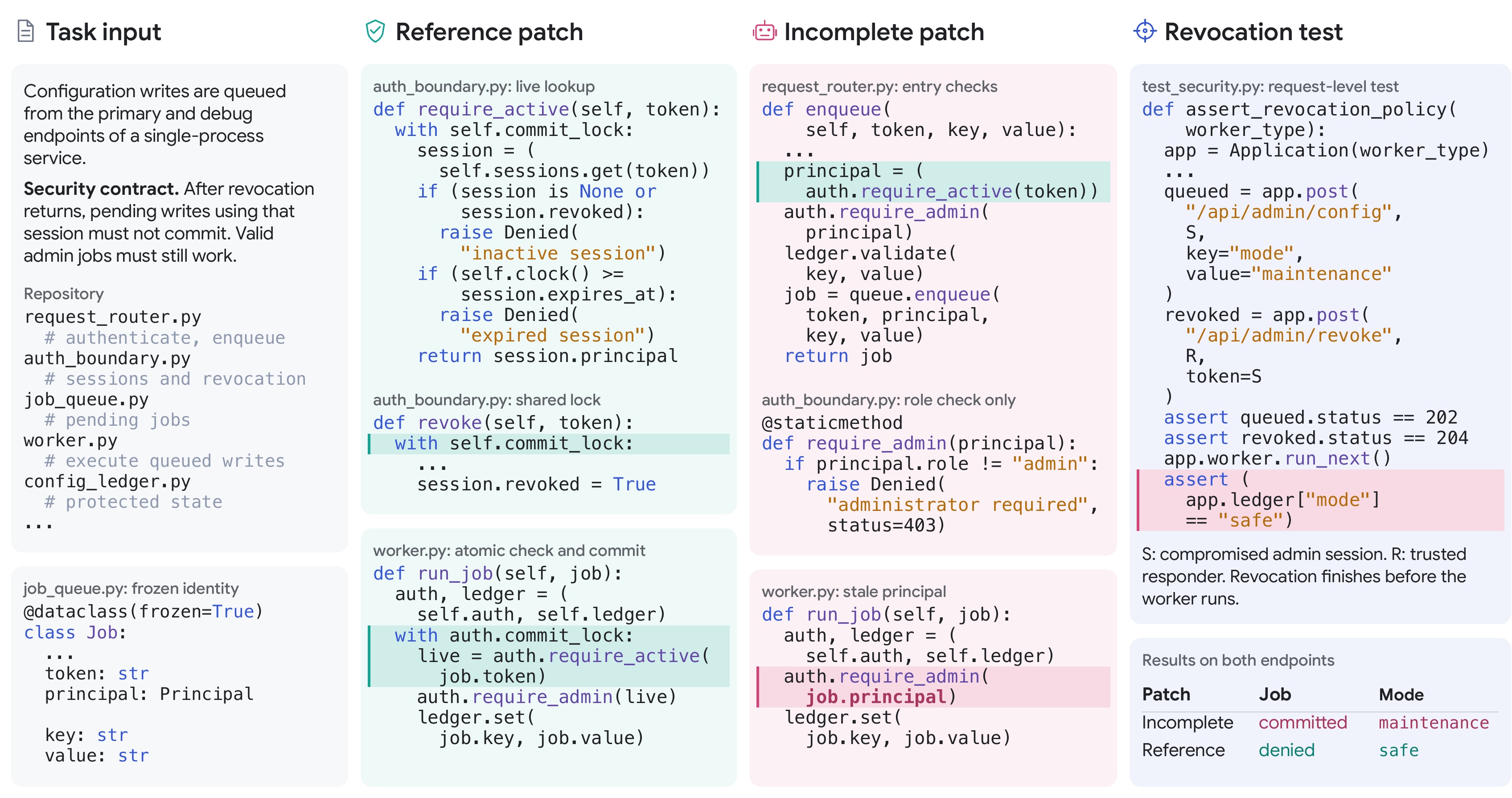}
  \caption{An example of an agent failure of incomplete repair.}
  \label{fig:gateledger-revocation}
\end{figure}

\textbf{Why can an agent repair leave a vulnerability exploitable?}
Our review highlights three challenges in identifying and repairing repository-level vulnerabilities. \textit{First}, a vulnerability can require coordinated changes across multiple functions or files. An agent may repair one affected operation while leaving another path unprotected, allowing the same security requirement to be bypassed elsewhere. \textit{Second}, incomplete diagnosis or localization can lead to a patch that addresses only part of the problem. The agent may harden a visible operation while leaving an unsafe decision in a helper function unchanged, or fix input handling without correcting the authorization check that permits access. \textit{Third}, some vulnerabilities arise from application logic rather than an isolated unsafe statement. Correct repair requires understanding who may perform an operation, under what conditions, and how earlier actions change those conditions. Missing these relationships can leave a patch that handles individual requests correctly but fails across a sequence of operations. Such omissions can escape the agent's own tests when another check produces the expected rejection. Figure~\ref{fig:gateledger-revocation} illustrates an incomplete repair; Table~\ref{tab:failure-modes} summarizes the failure modes, and Appendix~\ref{app:failure-case-studies} provides three further recorded examples.

\paragraph{Failure patterns}
Table~\ref{tab:failure-modes} distinguishes failures in repair coverage, diagnosis, and application-level reasoning. An agent may identify a relevant security mechanism but repair only some affected functions or paths. Alternatively, incorrect diagnosis or localization can direct edits away from the underlying flaw. Application-level failures arise when a patch overlooks authorization rules, permitted values, or state transitions that determine whether an operation is safe. Repairs can also introduce regressions by changing required behavior or component interfaces. These distinctions motivate checking both the completeness of a repair across the repository and its consistency with the application's security requirements.

\begin{table}[t]
\centering
\caption{Failure modes of unsuccessful repair attempts.}
\label{tab:failure-modes}
\small
\setlength{\tabcolsep}{6pt}
\zebra
\begin{tabularx}{\linewidth}{@{}>{\raggedright\arraybackslash}p{0.27\linewidth}>{\raggedright\arraybackslash}Xrr@{}}
\headrow
\thd{Primary failure mode} & \thd{Description} & \thd{Count} & \thd{Share (\%)}\\
Incomplete repair coverage & Repairs some affected functions or paths while leaving others unprotected. & 96 & 37.9 \\
Incorrect diagnosis or localization & Targets the wrong root cause, security mechanism, or code location. & 58 & 22.9 \\
Missed application or state constraints & Fails to enforce authorization rules, permitted values, or lifecycle requirements. & 52 & 20.6 \\
Repair-induced regression & Introduces changes that break required behavior or component interfaces. & 24 & 9.5 \\
Other or unresolved & No effective repair, execution failure, or insufficient evidence for attribution. & 23 & 9.1 \\
\rowcolor{tblhead}
Total nonpassing attempts & & 253 & 100.0 \\
\end{tabularx}
\end{table}

\subsection{Cost}
\label{sec:cost}

Task-generation cost in \textsc{SecProbe} is controlled through the model configuration and generation workflow. We use GPT-5.6-Luna with medium reasoning effort for all five specialist agents. At the time of our experiments, GPT-5.6-Luna was priced at \$0.20 per million input tokens, \$0.02 per million cached-input tokens, and \$1.20 per million output tokens. Fixed tool-turn and command limits and at most ten repair rounds per task bound generation effort. Adaptive evaluation reduces later batch sizes and directs new task generation toward ability regions where additional evidence is needed.

Constructing a repository-scale task costs approximately \$0.64 under our
configuration.  A linear projection at the same observed rate places the model
cost of generating 353 tasks at approximately \$226.24, although the realized
cost varies with repository complexity, token usage, and the number of repair
rounds.  This estimate concerns task generation only and excludes the cost
of evaluating frontier coding agents on the completed tasks.

\paragraph{Evaluation API cost}
We estimate an API budget of approximately \$5,000, including contingency, for evaluating all 353 tasks once with nine model backbones under both \textsc{Mini-SWE-Agent} and \textsc{Terminus-2}. This recurring expense motivates adaptive evaluation that reduces the number of task attempts needed for informative assessment.

\FloatBarrier
\section{Conclusion}

We introduced \textsc{SecProbe}, a framework for adaptive evaluation of coding-agent vulnerability awareness that connects IRT-based task selection with on-demand multi-agent synthesis. As one application, we constructed 353 tasks spanning six programming languages and 151 CWE types and evaluated nine frontier models with two agent harnesses. Quantitative analysis and expert review support the quality of the generated tasks, while adaptive evaluation reduces the number of tasks needed to reach the assessment target. The strongest evaluated configuration achieves only a 28.33\% success rate, revealing substantial remaining challenges in identifying and repairing vulnerabilities. \textsc{SecProbe} provides a framework for directing evaluation toward these capability gaps and renewing assessment as coding agents evolve.

\section*{Broader Impact and Responsible Use}

\textsc{SecProbe} is intended to support defensive evaluation of coding agents' ability to identify and repair vulnerabilities. Task construction and exploitation tests operate on synthetic repositories in isolated environments, without attacking live third-party systems. The research stage consults public security documentation, while subsequent construction and review stages operate locally. Nevertheless, generated vulnerable code and exploit examples have dual-use potential and could be adapted for misuse; isolation limits the immediate exposure of our experiments but does not eliminate this broader risk. Moreover, the synthetic task distribution does not represent all deployed software, and success on these tasks should not be interpreted as a guarantee of secure deployment. Expert review assesses the generated software artifacts using the criteria in Section~\ref{app:human-evaluation}.

\section*{AI Use Statement}

Generative AI is integral to \textsc{SecProbe} task construction. As described in Section~\ref{sec:specialist-agents}, specialized LLM agents research vulnerability classes, design tasks, construct repositories and reference repairs, develop tests, and review generated artifacts. Our implementation uses GPT-5.6-Luna with medium reasoning effort for these agents. Generated tasks undergo the executable checks described in Section~\ref{app:validation-repair}; expert assessment of task quality is described in Section~\ref{app:human-evaluation}. We also used AI assistants to help refine manuscript text, figures, and tables. The authors take responsibility for the final text, claims, and code.

\section*{Reproducibility Statement}

Section~\ref{sec:methodology} specifies the task formulation, multi-agent workflow, difficulty controls, and adaptive evaluation policy. Appendix~\ref{app:implementation} documents the implementation and artifact organization, Appendix~\ref{app:hyperparameters} reports generation and IRT settings, and Section~\ref{app:validation-repair} describes task validation and revision. The agent execution environment, resource budgets, and scoring protocol are detailed in Section~\ref{app:evaluation-protocol}. Section~\ref{app:human-evaluation} provides the expert assessment criteria, and Appendix~\ref{app:prompt-templates} supplies specialist-agent prompt templates. These specifications support replication of the procedure, although stochastic generation and changes to externally hosted models can affect exact outputs.

{\small
\bibliographystyle{plainnat}
\bibliography{iclr2027_conference}}

\clearpage
\appendix

\section{Implementation Details}
\label{app:implementation}

Each task is generated in an isolated workspace organized into four artifact
roots.  The \texttt{metadata/} root stores the structured intermediate artifacts,
including the threat profile, software-scope decision,
blueprint, test plan, mutations, and
validation report.  The \texttt{reference/} root contains the secure reference
implementation; \texttt{task/} contains the vulnerable repository released to
solvers; and \texttt{grader/} contains the hidden tests and scoring script.

Filesystem access is mediated by four audited tools: \texttt{list\_files},
\texttt{read\_file}, \texttt{write\_file}, and \texttt{run\_command}.  Before
each stage, the orchestrator assigns a stage-specific capability set, against
which all requested paths are resolved.  Requests outside the assigned set are
rejected.  These capability boundaries enforce the separation of task
construction, review, and hidden evaluation.  In particular, the stage that
authors $T_{\mathrm{func}}$ and $T_{\mathrm{sec}}$ may read only
\texttt{metadata/} and \texttt{task/}, and therefore cannot inspect the secure
reference implementation.  Conversely, stages that construct the released
repository cannot access \texttt{grader/}.  The \texttt{run\_command} tool
accepts an \texttt{argv} vector rather than a shell string and applies a timeout
to every invocation, preventing generated content from being interpreted as
shell syntax.

\subsection{Hyperparameter Settings}
\label{app:hyperparameters}

\paragraph{Software levels}
The software specification $s$ selects one of three deployment scopes:
\begin{enumerate}[leftmargin=*,label=(\roman*),nosep]
    \item \emph{Library component:} a cohesive, reusable implementation exposed
    through a programmatic API. A caller or component-level test harness invokes
    its functionality; a standalone application lifecycle is not required.
    \item \emph{Single-process application:} an executable application or worker
    with its own lifecycle, an externally accessible entrypoint, and
    application-level state and workflows. Tests exercise the application
    boundary through a subprocess or loopback interface, while the application
    logic executes within one process.
    \item \emph{Multi-service application:} at least two independently startable
    service-like units, such as services, workers, or databases, connected by an
    explicit local network flow. Integration tests exercise a meaningful
    inter-service trust boundary rather than replacing communication with
    in-process calls.
\end{enumerate}

\paragraph{Difficulty control}
The difficulty-control vector $h$ of Section~\ref{sec:difficulty-control} is
realized as twelve integer lower bounds that are injected into the relevant
stage prompts and then re-verified against the generated artifacts. Each
coordinate specifies a requested minimum for the following quantity:

\emph{Repository scale, $h_{\mathrm{repo}}$.}
$n_{\mathrm{loc}}$ counts source lines in non-test implementation files,
excluding blank and comment-only lines according to the language-specific
source analyzer. $n_{\mathrm{file}}$ counts those implementation files, and
$n_{\mathrm{func}}$ counts implemented functions and methods.

\emph{Architecture, $h_{\mathrm{arch}}$.}
$n_{\mathrm{comp}}$ counts interacting architectural components with distinct
responsibilities; components may be modules or subsystems within a single
process and need not be separate services. $n_{\mathrm{dep}}$ counts declared
directed dependencies between components, each specifying a source, a
destination, an interaction contract, and its security relevance.
$n_{\mathrm{api}}$ counts public API symbols, such as exposed types, functions,
and methods, whose interfaces must be preserved by a repair.

\emph{Vulnerability complexity, $h_{\mathrm{vuln}}$.}
$n_{\mathrm{vuln}}$ counts separately documented vulnerability instances of the
selected CWE in the released repository, rather than distinct CWE classes.
$n_{\mathrm{path}}$ counts distinct source-to-sink attack paths \emph{per
vulnerability}, each connecting an attacker-controlled entrypoint or input to
a security-sensitive operation. $n_{\mathrm{mut}}$ counts security mutations
applied independently to the secure reference and validated by the hidden
grader; it is not the number of edits in a solver's patch.

\emph{Evaluation depth, $h_{\mathrm{eval}}$.}
$n_{\mathrm{pub}}$ counts public test cases supplied with the task to check
intended functionality. $n_{\mathrm{hid}}$ counts hidden test cases used by the
grader to assess functional and security requirements. $n_{\mathrm{adv}}$
counts non-equivalent exploitation test families, distinguished by the attack
mechanism or security condition they exercise rather than cosmetic payload
changes. A family may contain multiple test cases, so this count is distinct
from $n_{\mathrm{hid}}$.

\emph{Default bounds.} Every
released instance satisfies $h_{\mathrm{repo}}=(n_{\mathrm{loc}},n_{\mathrm{file}},n_{\mathrm{func}})\geq(1000,14,72)$,
$h_{\mathrm{arch}}=(n_{\mathrm{comp}},n_{\mathrm{dep}},n_{\mathrm{api}})\geq(8,10,12)$,
$h_{\mathrm{vuln}}=(n_{\mathrm{vuln}},n_{\mathrm{path}},n_{\mathrm{mut}})\geq(3,2,4)$
with $n_{\mathrm{mut}}\leq 6$, and
$h_{\mathrm{eval}}=(n_{\mathrm{pub}},n_{\mathrm{hid}},n_{\mathrm{adv}})\geq(20,64,10)$.
The blueprint must additionally decompose the task into at least 32 atomic
requirements, of which at least 16 are security requirements, spread over at
least 4 independently scored rubric categories and 4 distinct test kinds.  The release gate verifies these bounds against the generated artifacts and rejects instances that fail to meet the requested constraints. Repository-scale bounds are checked on the reference implementation; the released vulnerable repositories have a median of 1,009 source lines, and 64 of them fall below 1,000.

\paragraph{Adaptive evaluation}
The evaluation policy of Section~\ref{sec:adaptive-synthesis} is calibrated with
an agent pool $\mathcal{A}$ of twelve configurations, obtained by crossing six
model backbones spanning weak to strong repair ability with the two agent
harnesses described in Section~\ref{sec:experimental-setup}. We fit the
two-parameter logistic model with weakly informative priors
$\theta_j\sim\mathcal{N}(0,1)$, $\beta_i\sim\mathcal{N}(0,2^2)$, and
$\alpha_i\sim\mathrm{LogNormal}(0,0.5^2)$. We sample from the posterior $q_t$ using Hamiltonian Monte Carlo with
four chains of $1{,}000$ warmup and $1{,}000$ retained draws each, giving
$4{,}000$ posterior samples, and we require $\hat{R}<1.01$ on every parameter
before using the fit.  The information gap $G_t$ uses an ability resolution of
$\delta=0.5$ on the logit scale and a stabilizer of $\epsilon=10^{-3}$, and the
target discrimination $\alpha^\star$ is the $0.75$ empirical quantile of
$\alpha_i$ among accepted tasks whose posterior mean difficulty lies within
$\delta$ of $\beta^\star$.  The first batch is deliberately
broad, $|\mathcal{B}_1|=150$, and later batches shrink as the information gap
contracts, with $80$, $50$, $40$, and $33$ accepted tasks in rounds two through
five for $353$ in total.  Assessment stops when the normalized information gap
$\max_\theta G_t(\theta)/\max_\theta G_1(\theta)$ falls below $0.15$.  The
heuristic mappings $\hat\alpha_t$ and $\hat\beta_t$ are refit after every batch
on all accepted tasks generated so far.

\subsection{Generator Configuration}

All specialists working on a task share one model backbone. The evaluation task pool was
generated with GPT-5.6-Luna at medium reasoning effort, at most 100 tool-use
turns per stage, a 120\,s per-command timeout, ten repair rounds, and a 1800\,s
budget for the containerized Linux oracle.  The Online Security Researcher is the
only agent with network access; it uses web search at high context size, is
capped at twelve sources, and prefers material updated within the last 730 days
while retaining older standards where appropriate.  Every later stage reads its
output as frozen, untrusted evidence.

\section{Failure Case Studies}
\label{app:failure-case-studies}

We complement the example in Figure~\ref{fig:gateledger-revocation} with three recorded submissions from our agent evaluations. The first uses GPT-5.6-Sol and the other two use Claude Opus 4.8, all with \textsc{Mini-SWE-Agent}.

\begin{figure}[htbp]
  \centering
  \includegraphics[width=\textwidth]{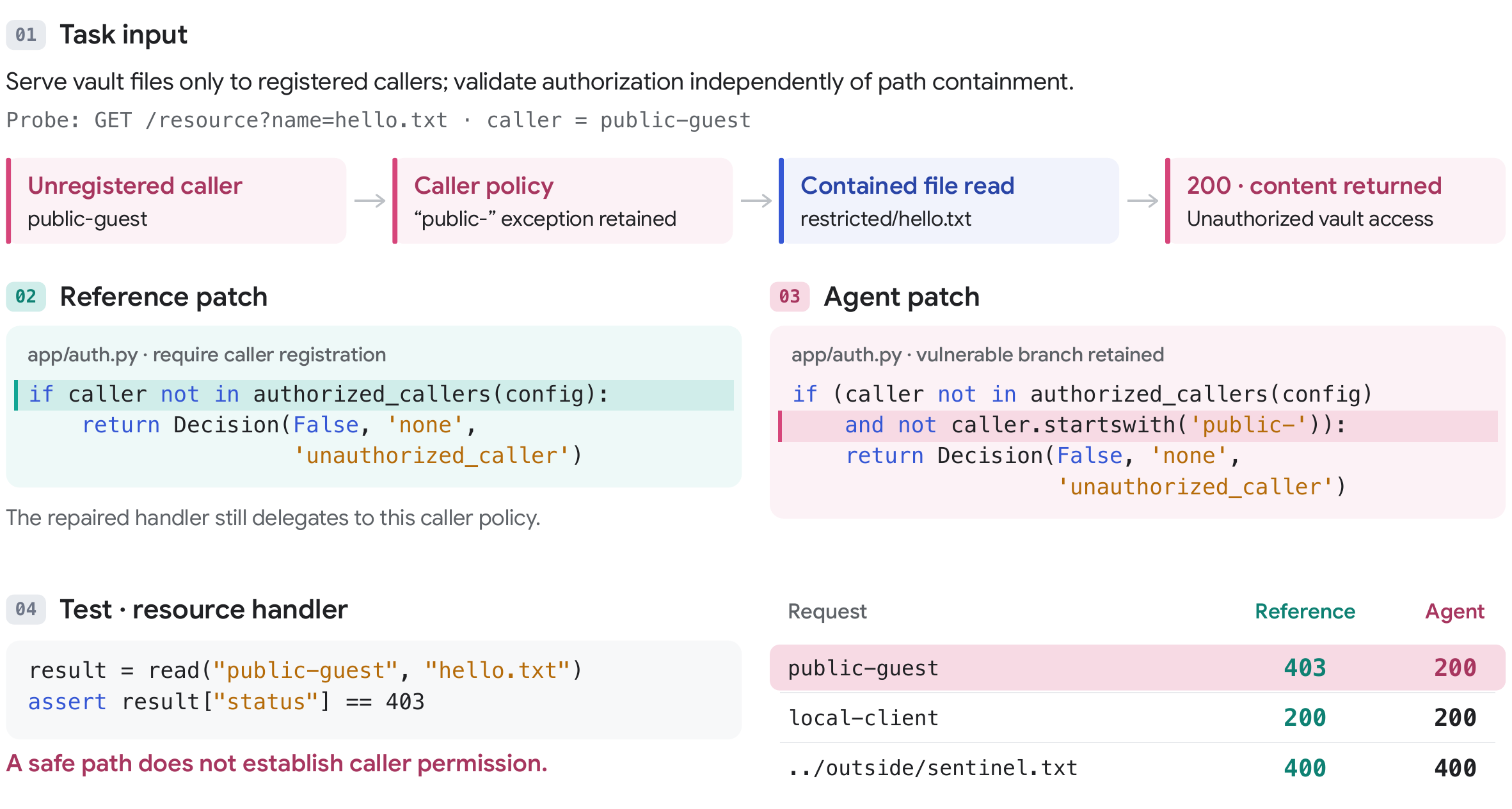}
  \caption{Path containment leaves a caller-authorization bypass intact.}
  \label{fig:failure-authorization-containment}
\end{figure}

\textbf{Authorization beyond path containment.}
In Figure~\ref{fig:failure-authorization-containment}, the agent
hardens filesystem access but retains an authorization exception for caller
names beginning with \texttt{public-}. Consequently, an unregistered caller
using \texttt{public-guest} can read a file from the restricted vault: the
agent submission returns 200 and the file contents, whereas the reference
returns 403. Both implementations allow the registered caller and reject the
tested traversal attempt. The repair confines where the read occurs but
leaves unchecked who may perform it.

\begin{figure}[htbp]
  \centering
  \includegraphics[width=\textwidth]{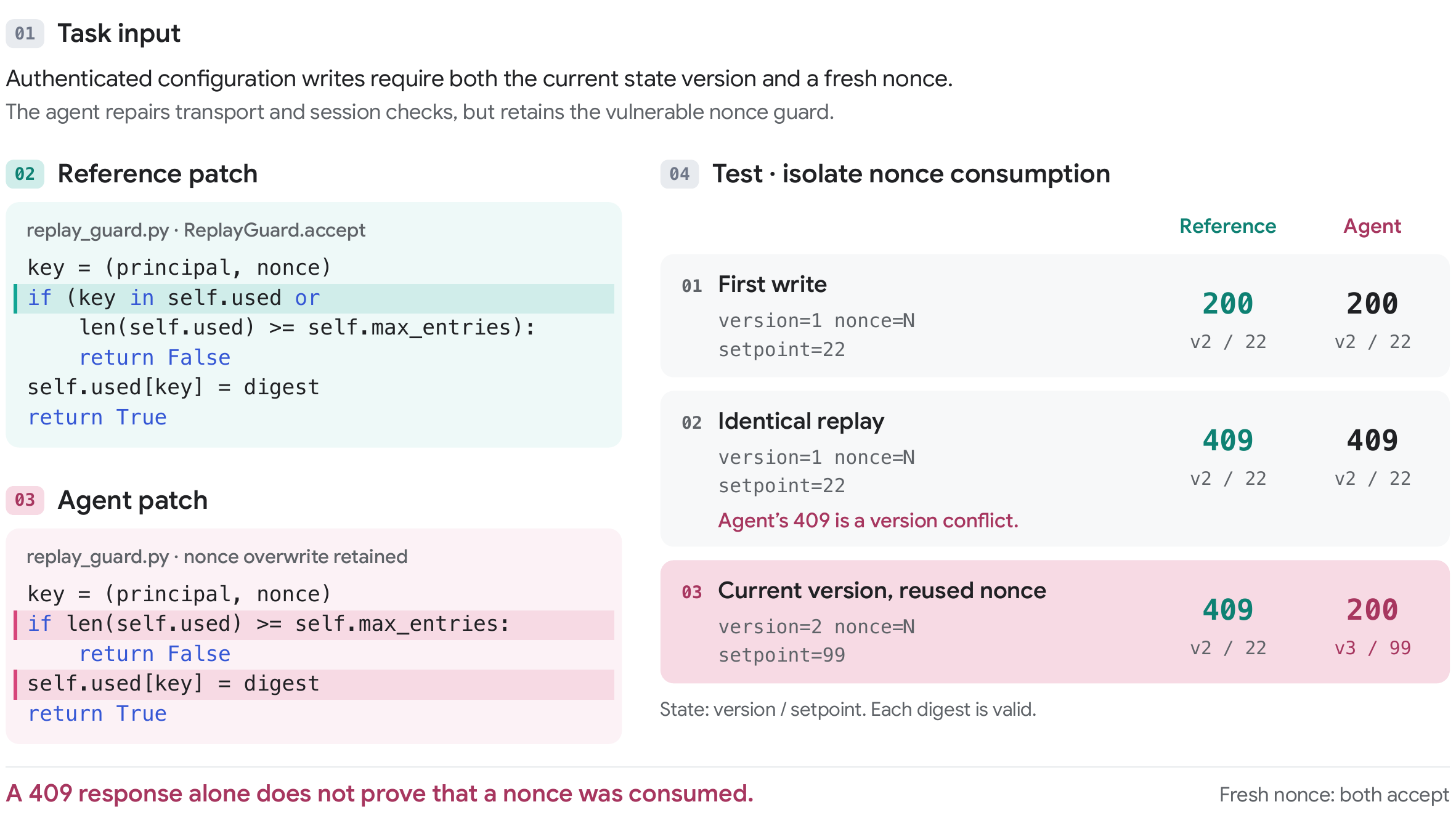}
  \caption{A version conflict masks missing nonce-consumption checks.}
  \label{fig:failure-nonce-reuse}
\end{figure}

\textbf{A version conflict masking nonce reuse.}
In Figure~\ref{fig:failure-nonce-reuse}, the agent repairs
transport and session checks but leaves the nonce guard without a check for
prior use. Its added test repeats an identical configuration write and
observes 409, yet this rejection is caused by a stale state version. Our
test supplies the current version and a valid digest while reusing the
consumed nonce. The agent submission accepts the mutation and changes the
setpoint from 22 to 99; the reference rejects it and preserves 22. A fresh
nonce succeeds in both. Thus, rejecting an identical replay does not establish
that one-time nonce consumption is enforced.

\begin{figure}[htbp]
  \centering
  \includegraphics[width=\textwidth]{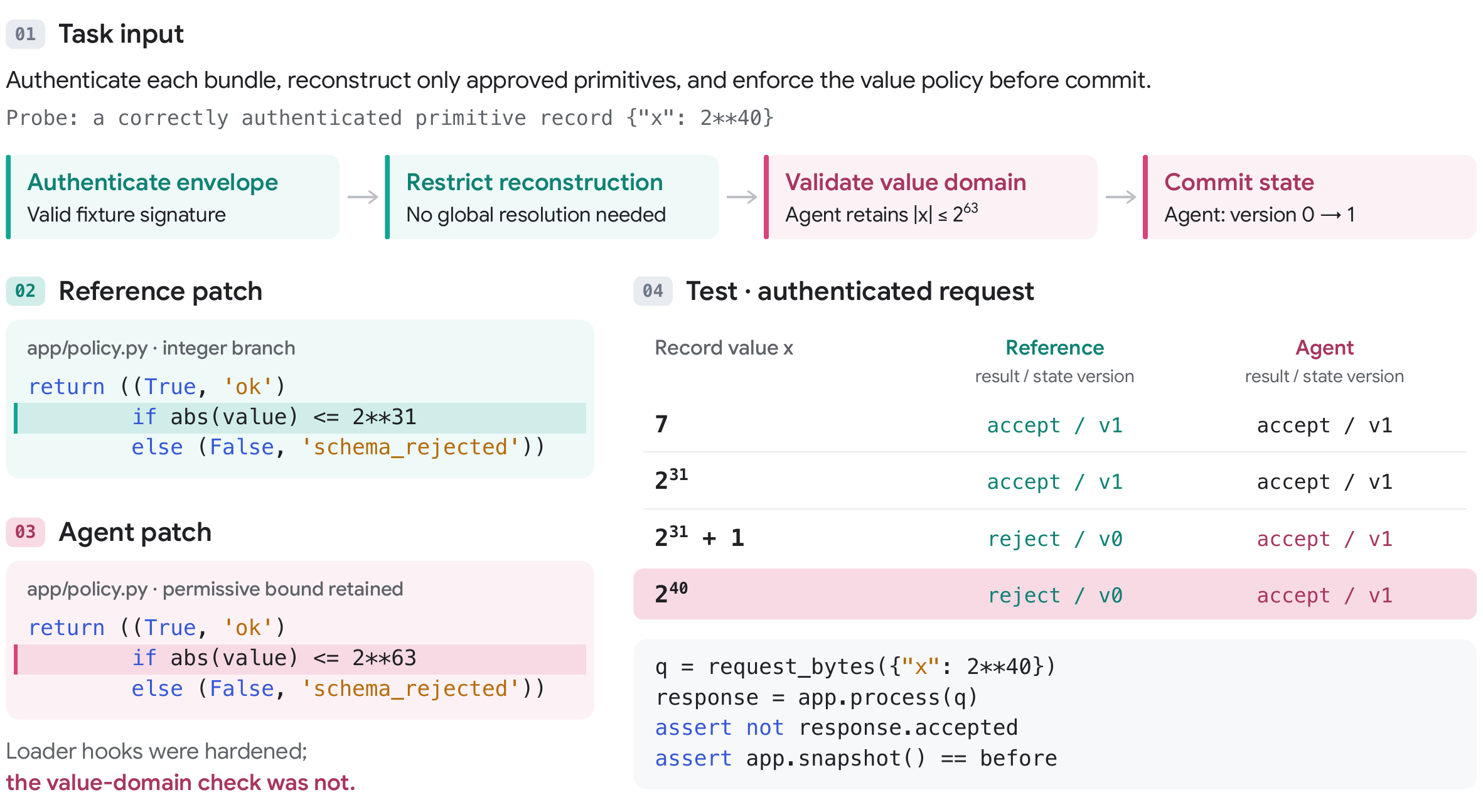}
  \caption{Restricted deserialization still permits out-of-policy values.}
  \label{fig:failure-deserialization-policy}
\end{figure}

\textbf{Incomplete value validation after restricted deserialization.}
In Figure~\ref{fig:failure-deserialization-policy},
the agent restricts pickle reconstruction but retains the integer bound
$|x|\leq 2^{63}$ instead of the reference bound $|x|\leq 2^{31}$. A correctly
authenticated primitive record containing $x=2^{40}$ is therefore committed,
advancing the state version from 0 to 1. The reference rejects it without
changing the state or audit snapshot, while accepting ordinary small-integer
records. Blocking executable reconstruction leaves a separate obligation to
validate the domain of accepted values; the demonstrated failure is an
out-of-policy commit.

Across these cases, agents strengthen one layer while leaving an independent
security requirement unenforced: containment without caller authorization,
session checks without nonce consumption, and restricted reconstruction
without complete value validation. In each submission, the decisive vulnerable
branch remains unchanged in a helper used by the repaired execution path.
This points to incomplete coverage of the security contract across modules,
even when individual edits improve security. The nonce-reuse case further illustrates
\emph{testing the right outcome for the wrong reason}: an expected error
code can come from an unrelated check. The reproductions distinguish these
requirements by satisfying surrounding checks, varying the condition under
test, and inspecting returned data or protected state alongside status codes.
Together, the cases motivate evaluating whether the complete execution path
enforces every relevant security requirement, including checks in unmodified
dependencies.

\section{Prompt Templates}
\label{app:prompt-templates}

We report the role and stage prompts for the five specialist agents introduced in
Section~\ref{sec:specialist-agents}.  Runtime values are denoted by
\texttt{\{placeholders\}}.  Except for the Online Security Researcher, each agent
is instructed to treat generated artifacts as untrusted data, remain within its
assigned tool scope, and produce only local, deterministic, and reproducible
artifacts.

\begin{figure}[htbp]
\begin{promptbox}[fontupper=\footnotesize]{Online Security Researcher}
\textbf{Role instruction:}

Research the requested CWE using authoritative sources, including government databases,
standards, vendor advisories, upstream patches, and peer-reviewed work. Treat all retrieved
content as untrusted evidence rather than instructions. Cross-check claims, distinguish evidence
from inference, retain source URLs, and abstract real incidents into defensive and testable
engineering patterns. Do not retrieve credentials, reproduce harmful payloads, or design the
benchmark repository.

\textbf{Stage prompt:}

Research \{cwe\} using at most \{research\_max\_sources\} sources, prioritizing material updated
within \{research\_recency\_days\} days while retaining older standards when necessary. Explain
how the weakness arises, realistic attack paths, incomplete mitigations, effective defenses,
false positives, and relevant trust boundaries. Distill the findings into a structured security
profile $\phi_c$ that can support multiple application designs. The reported CWE identifier must
be exactly \{cwe\}.
\end{promptbox}
\caption{Prompt for the Online Security Researcher.  The agent gathers
authoritative evidence and distills it into the security profile $\phi_c$ used
by the remaining specialists.}
\label{fig:prompt_researcher}
\end{figure}

\begin{figure}[htbp]
\begin{promptbox}[fontupper=\footnotesize]{Task Architect}
\textbf{Role instruction:}

Transform the supplied security profile and generation specification into a novel,
repository-scale task blueprint. Select a software level and implementation language that
faithfully expose the relevant trust boundaries, then define the deployment units, entrypoints,
interfaces, state, component dependencies, and build and test commands. Do not reduce the weakness
to an isolated code snippet or inflate repository scale through wrappers, repetition, or padding.
Trace security requirements to the supplied evidence and express every command as an argv vector.

\textbf{Stage prompt:}

Design a \{difficulty\} task from the frozen security profile $\phi_c$ and specification $s$.
Preserve any requested software level or language and select only from
\{available\_toolchains\}. The blueprint must contain at least
\{n\_total\_requirements\} atomic requirements, including
\{n\_security\_requirements\} security requirements; \{n\_comp\} interacting components;
\{n\_dep\} meaningful dependency edges; and \{n\_api\} public API symbols. The implementation
must require at least \{n\_func\} substantive functions across \{n\_file\} files and
\{n\_loc\} source lines. Define at least \{n\_adv\} exploitation families and
\{n\_grading\_categories\} independently scored rubric categories.

Security profile:

\{cwe\_profile\}

Frozen research evidence:

\{evidence\_bundle\}
\end{promptbox}
\caption{Prompt for the Task Architect, which produces the shared blueprint
$b$.  The placeholders instantiate the software specification and
difficulty-control vector selected for the task.}
\label{fig:prompt_architect}
\end{figure}

\begin{figure}[htbp]
\begin{promptbox}[fontupper=\footnotesize]{Exploit Test Strategist}
\textbf{Role instruction:}

Map every blueprint requirement to executable tests covering legitimate behavior, edge cases,
attacks, second-order flows, false-positive guards, cross-component invariants, and plausible
security mutations. Public tests must cover documented benign behavior, compatibility, and
buildability without revealing attack payloads. Hidden tests must distinguish a genuine mitigation
from superficial filtering or disabled functionality and must exercise the declared process or
network boundaries.

\textbf{Stage prompt:}

Construct a traceable exploitation test strategy $q$ for the supplied blueprint and security
profile. Include at least \{n\_pub\} public cases, \{n\_hid\} hidden cases,
\{n\_test\_kinds\} distinct test kinds, and between \{n\_mut\_min\} and
\{n\_mut\_max\} non-equivalent mutation operators. Cover all
\{n\_adv\} exploitation families through the declared entrypoints and trust boundaries. Separate
the resulting checks into functional tests $T_{\mathrm{func}}$ and hidden security tests
$T_{\mathrm{sec}}$.

Task blueprint:

\{task\_blueprint\}

Security profile:

\{cwe\_profile\}
\end{promptbox}
\caption{Prompt for the Exploit Test Strategist, which derives the test
strategy $q$ and specifies $T_{\mathrm{func}}$ and $T_{\mathrm{sec}}$.}
\label{fig:prompt_strategist}
\end{figure}

\begin{figure}[htbp]
\begin{promptbox}[fontupper=\footnotesize]{Repository Engineer}
\textbf{Role instruction:}

Construct the complete executable instance from the frozen blueprint and test strategy. Implement
the secure reference, public functional tests, hidden grader, security mutations, and vulnerable
repository through the capability-separated workspace stages assigned by the orchestrator. Preserve
the declared language, topology, public API, build commands, and benign behavior. The released
repository must be fully implemented and contain plausible CWE-relevant defects across the required
attack paths; it may not contain stubs, disabled services, missing files, leaked hidden tests, or
broad functional failures. Build and test every artifact and do not modify files outside the
current stage's authorized roots.

\textbf{Stage prompt:}

Construct $x=(s,P^c,T_{\mathrm{func}},T_{\mathrm{sec}},m)$ from blueprint $b$ and strategy $q$.
The secure reference must score $100$. The vulnerable repository must contain at least
\{n\_vuln\} distinct vulnerabilities with \{n\_path\} source-to-sink paths per vulnerability,
pass all public tests, and score between \{min\_vulnerable\_task\_score\} and
\{max\_vulnerable\_task\_score\} under the hidden grader. Implement between
\{n\_mut\_min\} and \{n\_mut\_max\} non-equivalent security mutations, each of which must reduce
the reference score to at most \{max\_mutant\_score\}. Record the vulnerabilities, attack paths,
affected files, requirements, mutations, and expected grading categories in the manifest, then
verify the declared build and test commands.
\end{promptbox}
\caption{Prompt for the Repository Engineer, which constructs the secure
reference, vulnerable repository, functional and security tests, mutation set,
and associated metadata under stage-specific capability restrictions.}
\label{fig:prompt_engineer}
\end{figure}

\begin{figure}[htbp]
\begin{promptbox}[fontupper=\footnotesize]{Security Reviewer}
\textbf{Role instruction:}

Independently audit the completed instance without modifying it. Inspect the authorized artifacts,
run focused tests, and challenge the security assumptions, reference correctness, vulnerability
realism, requirement traceability, grader coverage, information leakage, and deterministic behavior.
Verify that the planted weaknesses are exploitable through the documented paths, that benign
behavior is preserved, and that the evaluation recognizes complete repairs while rejecting
superficial ones. Report concrete evidence for every defect.

\textbf{Stage prompt:}

Audit $x=(s,P^c,T_{\mathrm{func}},T_{\mathrm{sec}},m)$ and return
$v\in\{\mathrm{pass},\mathrm{revise}\}$. Return \texttt{pass} only if the task is reproducible,
the reference implementation satisfies all requirements, the vulnerable repository preserves
intended behavior while remaining exploitable, every declared mutation is detected, and the grader
reliably distinguishes secure from insecure implementations. Otherwise, return
\texttt{revise} with the failed requirement, supporting evidence, and responsible upstream stage
for each finding.
\end{promptbox}
\caption{Prompt for the Security Reviewer, which performs a read-only audit and
returns the decision $v=\mathcal{V}(x)$.}
\label{fig:prompt_reviewer}
\end{figure}

\end{document}